\documentclass[aps,prd,twocolumn,showpacs,superscriptaddress,groupedaddress,nofootinbib]{revtex4}  

\usepackage{graphicx}
\usepackage{dcolumn}
\usepackage{bm}

\usepackage{orcidlink}

\usepackage{graphicx}	
\usepackage{amsmath}	
\usepackage{amssymb}	

\usepackage{epsfig,amsmath,natbib}
\usepackage{color,subfigure}
\usepackage{xcolor}
\usepackage[flushleft]{threeparttable}

\usepackage{mathrsfs,amssymb,amstext}
\usepackage{ulem}
\usepackage{url}
\usepackage{bm}
\usepackage{adjustbox}   
\usepackage{enumerate}

\begin{document}

\preprint{APS/123-QED}

\title{Primordial black holes
 and the velocity acoustic oscillations features in 21 cm signals from the cosmic Dark Ages}


  
\author{Zhihe Zhang\,\orcidlink{0009-0009-1496-1578}}
\affiliation{State Key Laboratory of Radio Astronomy and Technology, National Astronomical Observatories, CAS, 20A Datun
Road, Chaoyang District, Beijing 100101, P. R. China}
\affiliation{School of Astronomy and Space Science, University of Chinese Academy of Sciences, Beijing 100049, P. R. China}

\author{Bin Yue\,\orcidlink{0000-0002-7829-1181}}
\email{yuebin@nao.cas.cn}
\affiliation{State Key Laboratory of Radio Astronomy and Technology, National Astronomical Observatories, CAS, 20A Datun
Road, Chaoyang District, Beijing 100101, P. R. China}

\author{Yidong Xu\,\orcidlink{0000-0003-3224-4125}}
\affiliation{State Key Laboratory of Radio Astronomy and Technology, National Astronomical Observatories, CAS, 20A Datun
Road, Chaoyang District, Beijing 100101, P. R. China}

\author{Yin-Zhe Ma\,\orcidlink{0000-0001-8108-0986}}
\affiliation{Department of Physics, Stellenbosch University, Matieland 7602, South Africa}
\affiliation{National Institute for Theoretical and Computational Sciences (NITheCS), Stellenbosch University, Matieland 7602, South Africa}

\author{Xuelei Chen\,\orcidlink{0000-0001-6475-8863}}
\affiliation{State Key Laboratory of Radio Astronomy and Technology, National Astronomical Observatories, CAS, 20A Datun
Road, Chaoyang District, Beijing 100101, P. R. China}
\affiliation{School of Astronomy and Space Science, University of Chinese Academy of Sciences, Beijing 100049, P. R. China}


\begin{abstract}
{
Astrophysical luminous objects such as the first stars have not yet formed in the Dark Ages. However, primordial black holes (PBHs) always exist throughout cosmic history since the inflation epoch. During the Dark Ages, PBHs may accrete the ambient gas and release radiation like astrophysical luminous objects, change the cosmic radiation field, the thermal status of the intergalactic medium (IGM), and the hydrogen spin temperature. The accretion rate is modulated by the relic supersonic relative streaming velocities between dark matter (DM) and baryons, imprinting Velocity Acoustic Oscillations (VAOs) features in the 21 cm power spectrum. Such VAOs features could be a promising probe for detecting the PBHs in Dark Ages. We find that even if PBHs comprise only a small fraction of DM, they can generate VAOs wiggles with a relative amplitude up to $\sim 30\%$ in Dark Ages. For example, for PBHs with mass $M_{\rm PBH,rec}=200~M_\odot$ and mass fraction in the total DM $f_{\rm PBH,rec}\sim 10^{-13}$ at the recombination era,
VAOs features appear  at $z\sim 20$; if $f_{\rm PBH,rec}\sim 3\times 10^{-10}$, then VAOs features could appear as early as $z\sim 40$. Moreover, the redshift evolution of the VAOs features exhibits clearly separated stages dominated by inhomogeneous Ly$\alpha$ scattering, and inhomogeneous X-ray heating, respectively. It reflects the characteristics of PBHs (mass and fraction in total DM) and their interactions with the IGM.  We also estimate that, the VAOs wiggles at $z\sim20$ are detectable for the upcoming SKA-low AA*, while wiggles at $z\sim 40$ are detectable for an hypothetic lunar surface-based interferometer array in the future.}
\end{abstract}

\maketitle

\section{Introduction}\label{sec:intro} 
 
The cosmic Dark Ages starts from $z\sim 1100$ when protons and electrons combine into neutral Hydrogen (H) atoms; ends  at $z\sim 50 -20$ when a massive number of first stars form and radiate the first light that heavily changes the cosmic radiation field, the intergalactic medium (IGM) thermal status, and the H spin temperature \cite{2006PhR...433..181F,2012RPPh...75h6901P}. Dark Ages is generally considered as an ideal stage for probing primordial fluctuations, as the H spin temperature is purely determined by scattering with CMB photons and collisions with other particles. The 21 cm signal has not yet been contaminated by astrophysical sources, performing as a good tracer of gas density field \cite{2021ExA....51.1641K,2022JCAP...06..020F}. However, in Dark Ages there might be  other luminous objects that can change the 21 cm signal.

Primordial Black Holes (PBHs) arise from small-scale primordial curvature fluctuations and form in the early Universe, much earlier than the first stars \cite{2022JCAP...03..012C,2021JCAP...05..051C,2023ApJ...949...64C}.  Their mass ranges from the Planck mass  to $\sim 10^{5}~M_{\odot}$ \cite{2005astro.ph.11743C,2005PhRvD..71j4009H}. PBHs with mass   $\lesssim10^{15} ~\rm g$ would have evaporated through Hawking radiation \cite{1979Natur.278..605C,2010PhRvD..81j4019C}, whereas stellar-mass and intermediate-mass PBHs can emit radiation by accreting surrounding gas \cite{2017PhRvD..95d3534A,Ricotti2007_lambda,2008ApJ...680..829R}, and may even launch powerful relativistic jets that generate strong radio emission \cite{2303.06616,2022MNRAS.517L...1T,2020A&A...638A.132B}. Even below the currently allowed upper limits set by observations \cite{2016PhRvL.116t1301B,2012PhRvD..86d3001B,2017PhRvD..96h3524P,2020PhRvR...2b3204S,2021JPhG...48d3001G,2021JCAP...05..051C,2023ApJ...949...64C}, their number density can still far exceed that of the black hole seeds formed from first stars. As a result, PBHs can produce strong radiation that affects the cosmic radiation field, the thermal and ionization states of the IGM in Dark Ages \cite{2017PhRvD..96h3524P,2021MNRAS.508.5709Y,2008arXiv0805.1531M,2017PhRvD..95h3006C}, and even alter the CMB anisotropies \cite{2021PhRvD.104f3534J,2025PhRvD.111d3505J}.

Dark Matter (DM) decouples from radiation much earlier than recombination era  and evolves freely since then, while baryon remains coupled to radiation until recombination era. The interaction between gravity and radiation pressure generates large-scale density shift and relative streaming motion between DM and baryon.  At recombination era, the root-mean-square (rms) amplitude of these relative streaming velocities is $\sim 30~\rm km\ s^{-1}$ \cite{2010PhRvD..82h3520T,2014IJMPD..2330017F}. The relative streaming velocity coherently oscillates, with smallest coherent comoving scale $\sim 3$ Mpc \cite{2010PhRvD..82h3520T}. 
After recombination era, such spatial structure pattern is frozen, so that the corresponding oscillatory features remain imprinted in the power spectrum of the large-scale streaming velocity field. These features are commonly referred to as velocity-induced acoustic oscillations (VAOs) \cite{2010PhRvD..82h3520T,2014IJMPD..2330017F,2023PhRvD.107b3524S, 2022PhRvD.106f3504F, 2022PhRvD.105j3529S} and the relative streaming velocity decays with the Universe expands, approximately as $\sigma_{\rm rms}\approx 30(1+z)/1100~\rm km~s^{-1}$ \cite{2014PhRvD..89b3519D}. At Cosmic Dawn, such velocity is still supersonic and significantly reduces the first stars formation in minihalos \cite{Greif2011ApJ,2011MNRAS.418..906T,Stacy2011ApJ,Fialkov2012MNRAS,2023MNRAS.525..428H,Conaboy2023MNRAS}, while may enhance the formation of black holes \citep{Hirano2017Sci,Kimura2025ApJ}, leaving imprints in the 21 cm signal \cite{2010JCAP...11..007D,McQuinn2012ApJ,2012Natur.487...70V,Bovy2013ApJ,2013MNRAS.432.2909F,2014PhRvD..89h3506A,Cohen2016MNRAS,2019PhRvD.100f3538M,2022MNRAS.511.3657M,2023ApJ...950...20S}. The VAOs features could be a probe of both first stars, small-scale structures  and cosmology \cite{ZhangXin_VAO,2022MNRAS.511.3657M,2023MNRAS.525..428H,Sarkar2023PhRvD}.

The PBHs radiation is tightly linked to their accretion properties \cite{Merloni2003,Lusso2012MNRAS,2303.06616}. The accretion rate is strongly affected by the motion of the PBHs relative to the surrounding gas \cite{Bondi1952,2008ApJ...680..829R}. Since PBHs comprise DM, this motion must be modulated by the DM-baryon relative streaming velocity field \cite{2008ApJ...680..829R,2017PhRvD..95d3534A}. Such an effect can leave oscillating features in the cosmic free electron fraction field, and disturb the CMB power spectrum  \citep{2021PhRvD.104f3534J,2025PhRvD.111d3505J}.
It is therefore natural to ask whether radiation produced by PBHs accretion can further map such VAOs features into the power spectrum of the 21 cm signal in Dark Ages, through Ly$\alpha$ scattering and  X-ray heating, or even radio background. If so it provides a distinctive signature of PBHs-induced heating and coupling in the Universe stage before the formation of first stars.
 
Observing the 21 cm signal from Dark Ages is more challenging than Cosmic Dawn and epoch of reionization, because the frequency and amplitude of the signal are lower, while the foreground and noise level are higher. The upcoming low-frequency array of Square Kilometer Array (SKA-low) has the lowest frequency boundary of 50 MHz, and therefore has the ability to receive 21 cm signals from the Dark Ages with $z\lesssim 27$ \cite{2009IEEEP..97.1482D,2013ExA....36..235M,2015aska.confE...1K}. 
However, for signals from higher redshifts, the absorption and distortion effects by the ionosphere would significantly distorts the signals. One solution is to construct interferometer arrays at the farside of the Moon, where the ionosphere is negligible and the radio frequency interferences from the Earth is blocked. Some of such interferometers are the lunar-orbit DSL \cite{2021RSPTA.37990566C}, and lunar surface-based FARSIDE \cite{2019arXiv191108649B}, ALO \cite{2024AAS...24326401K} and the FarView \cite{2024AdSpR..74..528P}. 

In this paper, we investigate the impacts of the DM-baryon relative streaming motion on PBHs accretion, and their imprints on 21 cm signals. We restrict our investigation in Dark Ages when the radiation from any astrophysical objects, i.e. first stars,  galaxies and black holes form via astrophysical processes, is still negligible. In this case, only radiation from the PBHs heats and ionizes the IGM, and finally leaves its large-scale features in the 21 cm signal. This paper is organized as follows: In Sec. \ref{sec: v_db_accretion} we introduce the DM-baryon relative stream motion, and the influence on PBHs accretion rate. In Sec. \ref{sec:X-ray} we investigate the VAOs features in the IGM temperature field, arising from inhomogeneous X-ray heating. In Sec. \ref{sec:21cm} we investigate the VAOs features in 21 cm signals due to inhomogeneities in both X-ray heating and Ly$\alpha$ scattering. In Sec. \ref{sec:radio-bg} we investigate the influences if PBHs also produce strong radio emission that build a strong radio background at the Dark Ages. In Sec. \ref{sec:detectability} we estimate the detectability for SKA-low and a hypothetic lunar surface-based array. Finally we summarize the results and give discussions in Sec. \ref{sec:summary}.

\section{The  DM-baryon relative streaming velocities and the PBHs accretion}\label{sec: v_db_accretion}

\subsection{The DM-baryon relative streaming velocities}

Since PBHs comprise DM, we assume their large-scale motions trace DM. This is an approximation but it is reasonable, see some discussion about this point in Sec. \ref{sec:discussion}.
The relative motions between PBHs and baryon is described by the DM-baryon relative streaming velocities, 
\begin{equation}
    \vec{v}_{\rm db}=\vec{v}_{\rm d}-\vec{v}_{\rm b},
\end{equation}
where the DM velocity field $\vec{v}_{\rm d}$, the baryon velocity field $\vec{v}_{\rm b}$, and the corresponding density perturbations $\delta_{\rm d/b}$, are linked by the linearized continuity equations
\begin{equation}
    \dot{\delta}_{\rm d/b}+a^{-1}\nabla\cdot\vec{v}_{\rm d/b}=0, 
\end{equation}
where $a=1/(1+z)$ is the scale factor. 

In Fourier space the above equations are  
\begin{align}
    \vec{v}_{\rm d/b}(\vec{k},a)&=-ia\frac{\vec{k}}{k^2}\dot\delta_{\rm d/b}(\vec{k},a) \nonumber \\
    &=-ia\frac{\vec{k}}{k^2}\dot{\mathcal{T}}_{\rm d/b}(\vec{k},a)\delta_{\rm pri}(\vec{k})
\end{align}
where $\mathcal{T}_{\rm d/b}(\vec{k},a)$ are the transfer functions of the DM and baryon power spectrum respectively, $\delta_{\rm pri}(\vec{k})$ is the primordial density fluctuations.

Following the methods described in Ref. \cite{ZhangXin_VAO}, we generate the fields of $\vec{v}_{\rm d}$, $\delta_{\rm d}$, $\vec{v}_{\rm b}$ and $\delta_{\rm b}$ respectively. The fields have a box length of 1200 Mpc and resolution of 3 Mpc, which is the coherence scale of the DM-baryon relative streaming velocities \cite{2010PhRvD..82h3520T}. 

\begin{figure}
    \centering   
    {
\subfigure{\includegraphics[width=0.45\textwidth]{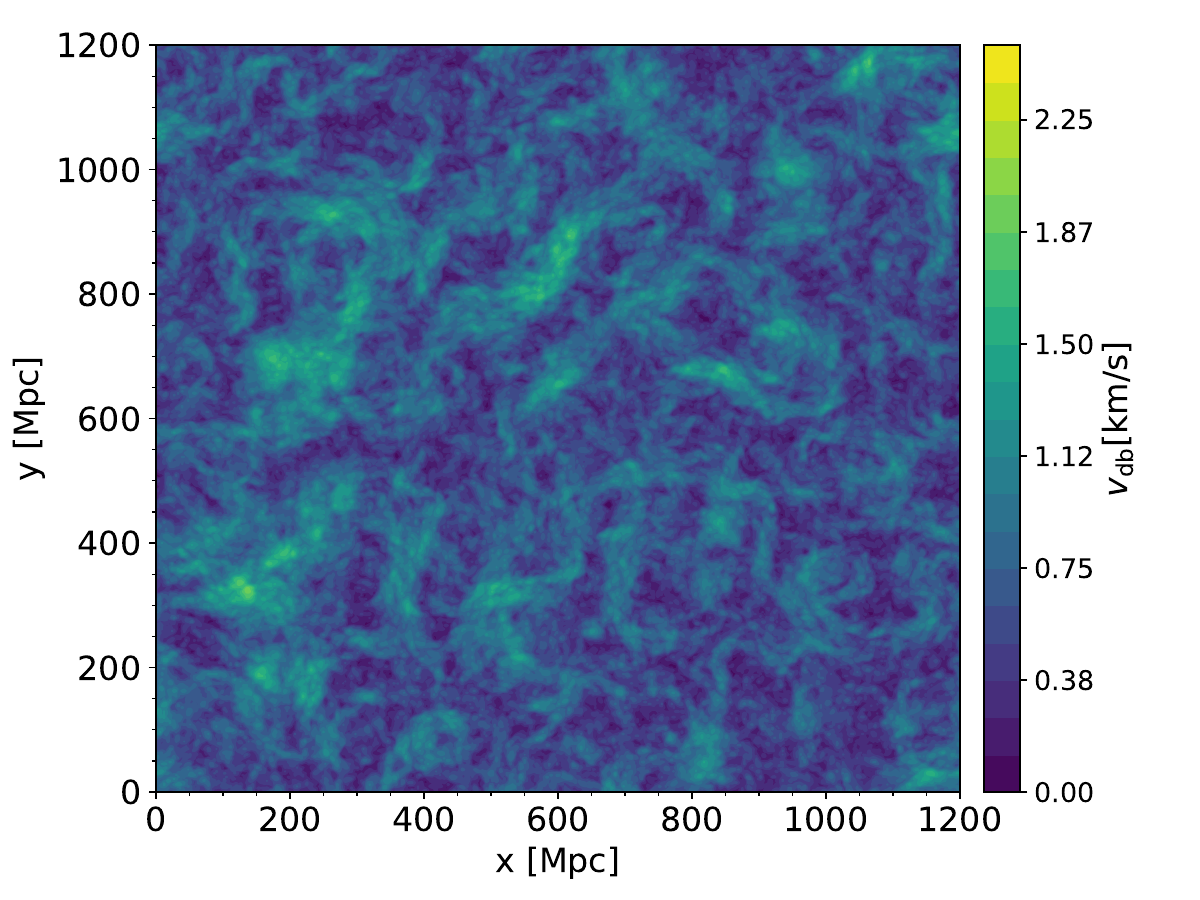}} 
\subfigure{\includegraphics[width=0.45\textwidth]{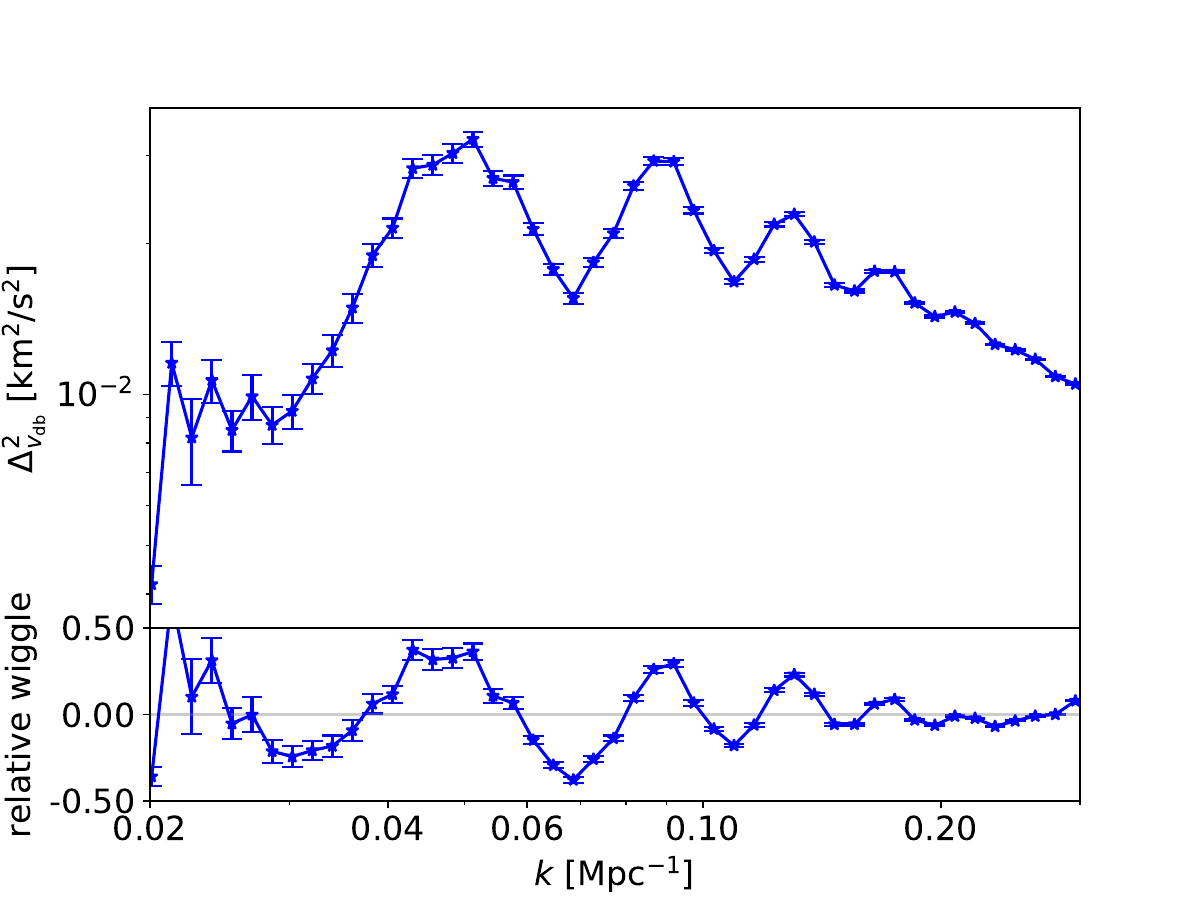}} 
\caption{
$Top$: A slice of the DM-baryon relative streaming velocity field at $z=20$. 
$Bottom$: The power spectrum of the DM-baryon relative streaming velocity field,
$\Delta_{v_{\rm db}}^2(k)=k^3/(2\pi^2)P_{v_{\rm db}}(k)$, at $z=20$ (top sub-panel). 
We fit the power spectrum by a 5th degree polynomial $\Delta^2_{v_{\rm db},\rm poly}(k)$, then show the relative amplitude of the VAOs wiggles, $[\Delta_{v_{\rm db}}^2(k) - \Delta^2_{v_{\rm db},\rm poly}(k)]/\Delta^2_{v_{\rm db},\rm poly}(k)$, in the bottom sub-panel.  
Throughout this paper, errorbars on the power spectrum curves and relative amplitude curves refer to the 1$\sigma$ sample variance of the signal calculated from the generated fields.
    \label{fig:v_bc}
   }
   }
\end{figure}

In Fig. \ref{fig:v_bc} we show a slice of the ${v}_{\rm db}$ field at $z=20$, and the corresponding power spectrum. The oscillation (wiggle) features are clearly seen in the power spectrum, with relative amplitude up to $\sim 40\%$, compared to a smooth reference described by a 5th degree polynomial.

\subsection{The PBH accretion and mass growth}

The accretion rate of a PBH is ~\cite{Bondi1952}
\begin{equation}
\dot{M}_{\rm B}=4 \pi \lambda {\rho}_{\rm b} v_{\rm eff} r^2_{\rm B,eff},
\label{eq:M_dot}
\end{equation}
where $\lambda$ is a dimensionless parameter describing the net accretion efficiency, $\rho_{\rm b}$ is the baryon density, $v_{\rm eff}$ represents the relative velocity between the PBH and the ambient gas, and $r_{\rm B,eff}$ denotes the effective Bondi radius.

Typically, $\lambda\sim10^{-2}-10^{-3}$ \cite{Perna_2003,Hektor2018PRD}, in some models it can reach $\sim 0.1-1$ \cite{2017PhRvD..95d3534A, 2020PhRvR...2b3204S,Ricotti2007_lambda}.
We adopt the parameter $\lambda=10^{-3}$ as the fiducial value in this paper.
For a PBH reside in a dark matter halo with mass $M_{\rm h}$, the effective Bondi radius $r_{\rm B,eff}$ is the solution of the equation
\begin{equation}
r_{\rm B,eff}=\frac{G(M_{\rm PBH}+M_r(<r_{\rm B,eff},M_{\rm h}))}{v_{\rm eff}^2}, \label{eq:rB}
\end{equation}
where $M_r(<r, M_{\rm h})$ denotes the enclosed mass within a radius $r$. The total dark matter halo mass \cite{Ricotti2007_lambda},
\begin{equation}
    M_{\rm h}=M_{\rm PBH}\left(\frac{3000}{1+z}\right).
    \label{eq:Mh}
\end{equation}
and radius
\begin{eqnarray}
r_{\rm h}&=& 0.339 r_{\rm turn} \nonumber \\
&=& 0.339\frac{58}{1+z}\left(\frac{M_{\rm h}}{ M_{\odot}}\right)^{\frac{1}{3}}\text{pc},
\end{eqnarray}
where $r_{\rm turn}$ is the turnaround radius.
If the dark matter halo has a power-law density profile and is truncated at the halo radius 
$r_{\rm h}$, say $\rho\propto r^{3-p}$ and take $ p=0.75$~\cite{Ricotti2007_lambda}, then
\begin{equation}
r_{\rm B,eff}\approx r_{\rm B,h}\left(\frac{r_{\rm B,eff}}{r_{\rm h}}\right)^p.
\end{equation}
The effective velocity $v_{\rm eff}$ is contributed from both the sound speed and the relative motion between the black hole and the ambient gas,
\begin{equation}
v_{\rm eff}=\sqrt{v_{\rm db}^2+c_{\rm s}^2}.
\label{eq:v_eff}
\end{equation}
We assume the PBH always trace the DM distribution, therefore the PBHs-baryon relative motion is just the  DM-baryon relative motion with speed $v_{\rm db}$.
 
During accretion, a PBH grows as 
\begin{equation}
    \frac{dM_{\rm PBH}}{dt} = (1-\epsilon_{\rm rad})\dot{M}_{\rm B},
    \label{eq:dM_PBH_dt}
\end{equation}
where $\epsilon_{\rm rad}$ is the radiative efficiency \cite{Hektor2018PRD},
\begin{equation}
\epsilon_{\rm rad}=0.1\times
\begin{cases} 
\left(  \frac{\dot{m}}{\dot{m}_{\rm crit}}   \right)^\beta~~~~~~\dot{m}\le \dot{m}_{\rm crit} \\
1~~~~~~~~~~~~~~~~\dot{m}> \dot{m}_{\rm crit}, 
\end{cases}
\end{equation}
where $\dot{m}=\dot{M}_{\rm B}/\dot{M}_{\rm Edd}$ is the dimensionless accretion rate, $\dot{m}_{\rm crit}=0.1$, $\dot{M}_{\rm Edd}=L_{\rm Edd}/c^2$, and $L_{\rm Edd}=1.26\times10^{38}({M_{\rm PBH}}/{M_{\odot}})\ \rm erg\ s^{-1}$ is the Eddington luminosity. We set the index $\beta=1$, and assume that  during the PBH growth, Eq. (\ref{eq:Mh}) always holds. The accretion rate before the recombination era is much smaller than the latter stages \citep{2008ApJ...680..829R}, so when integrating the Eq. (\ref{eq:dM_PBH_dt}), we start from the recombination era. The PBHs mass at this time is $M_{\rm PBH,rec}$. Throughout this paper, we define the mass fraction of PBHs in the total DM $f_{\rm PBH}\equiv\rho_{\rm PBH}/\rho_{\rm DM}$, where $\rho_{\rm PBH}$ and $\rho_{\rm DM}$ are cosmic PBHs density and cosmic DM density respectively. The value of this ratio at the recombination era, which is the initial condition of our differential equations, is $f_{\rm PBH,rec}$. This should not be confused with the ``initial PBHs abundance'' which generally refers to the ratio between the cosmic PBHs density and the sum of cosmic total matter and radiation density at the time when PBHs just formed \cite{2022Univ....8...66E,2018CQGra..35f3001S}.
See an explicit clarification in Sec. \ref{sec:discussion}.

\begin{figure}
    \centering   
    {
    \subfigure{\includegraphics[width=0.45\textwidth]{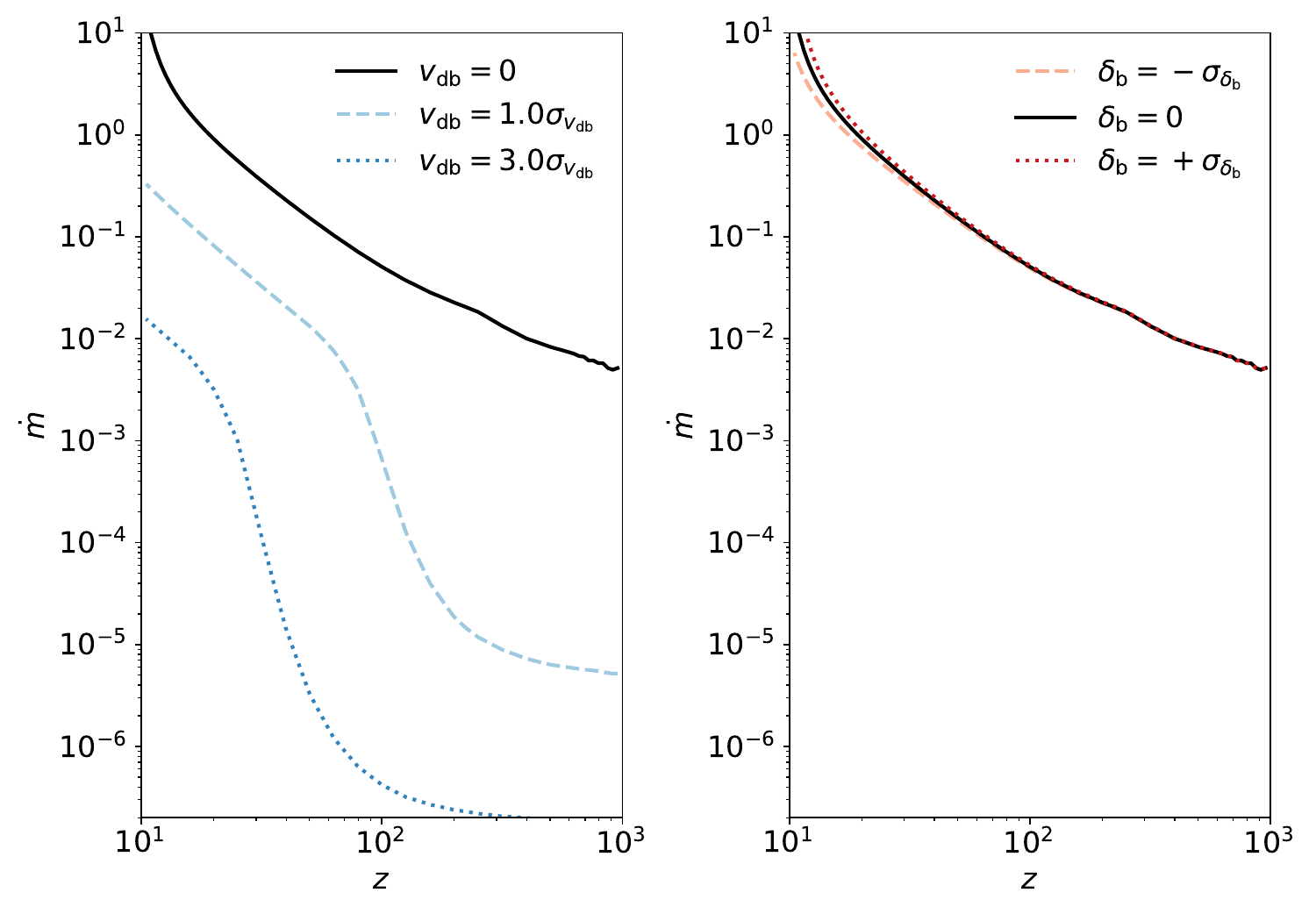}} 
    \caption{The redshift evolution of the dimensionless accretion rate for a PBH with $M_{\rm PBH,rec} = 200~M_{\odot}$. Both panels adopt $\delta_{\rm b}=0$ and $v_{\rm db}=0$ as the reference case (black solid line). $Left$: The $\dot{m}$ for varying $v_{\rm db}$. $Right$: The $\dot{m}$  for varying $\delta_{\rm b}$. Obviously, $\dot{m}$ is more sensitive to  $v_{\rm db}$ than $\delta_{\rm b}$.  
    }
    \label{fig:m_dot}
    }
\end{figure}

Fig. \ref{fig:m_dot} shows the evolution of the accretion rate for a PBH with $M_{\rm PBH,rec}=200~M_\odot$
in environment with different $v_{\rm db}$ or different $\delta_{\rm b}$. For this moment we ignore the heating of the PBH radiation to the IGM, which will be investigated in next section. We see that the PBH accretion rate is much more sensitive to the relative streaming velocities  than to density fluctuations. For $v_{\rm db}=\sigma_{v_{\rm db}}$, the accretion rate is reduced by more than 1 order compared to the $v_{\rm db}=0$; however, for $\delta_{\rm b}=\pm \sigma_{\delta_{\rm b}}$, the accretion rate just changes slightly. This indicates that the spatial variations of accretion rates are  mainly controlled by $v_{\rm db}$, giving rise to the characteristic VAOs features on the power spectrum of $\dot{m}$.

\begin{figure}
\centering
\subfigure{\includegraphics[width=0.45\textwidth]{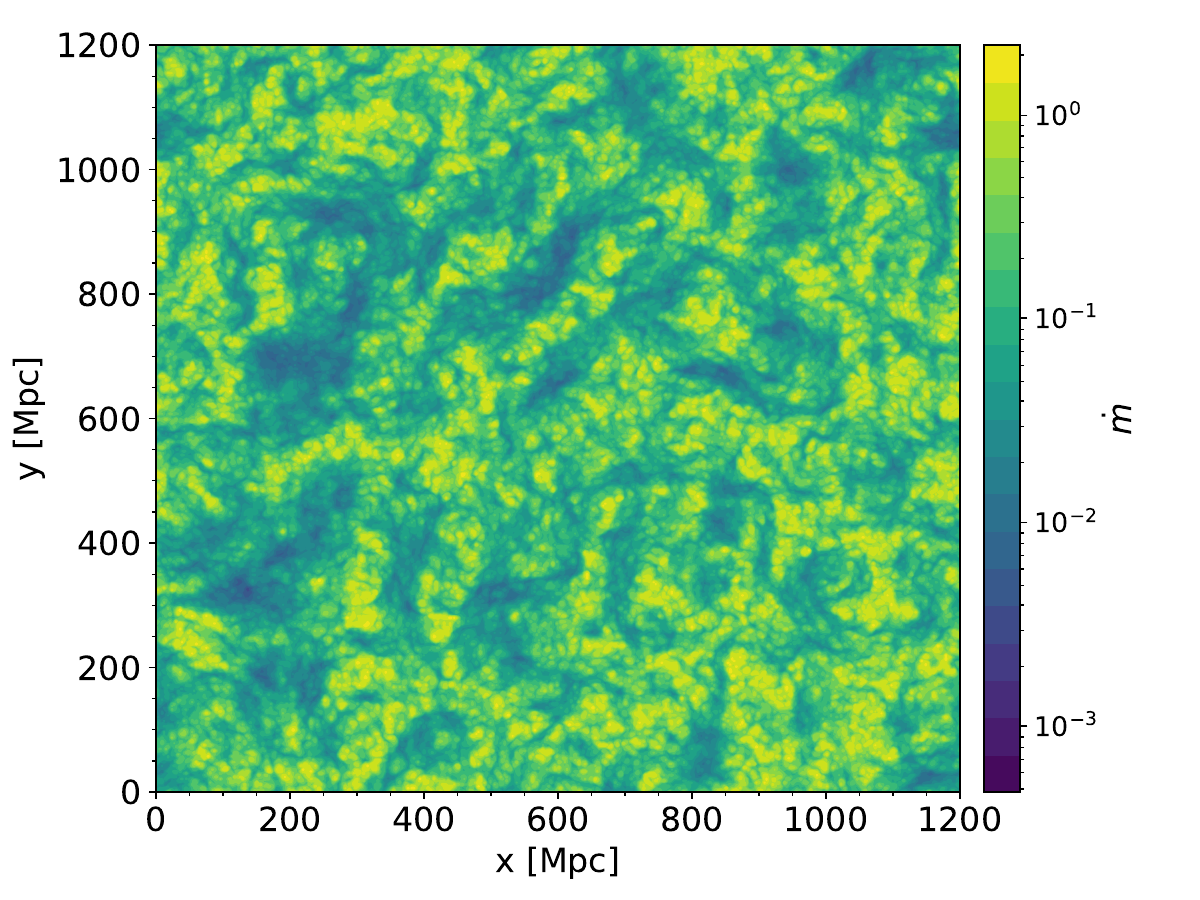}}
\subfigure{\includegraphics[width=0.45\textwidth]{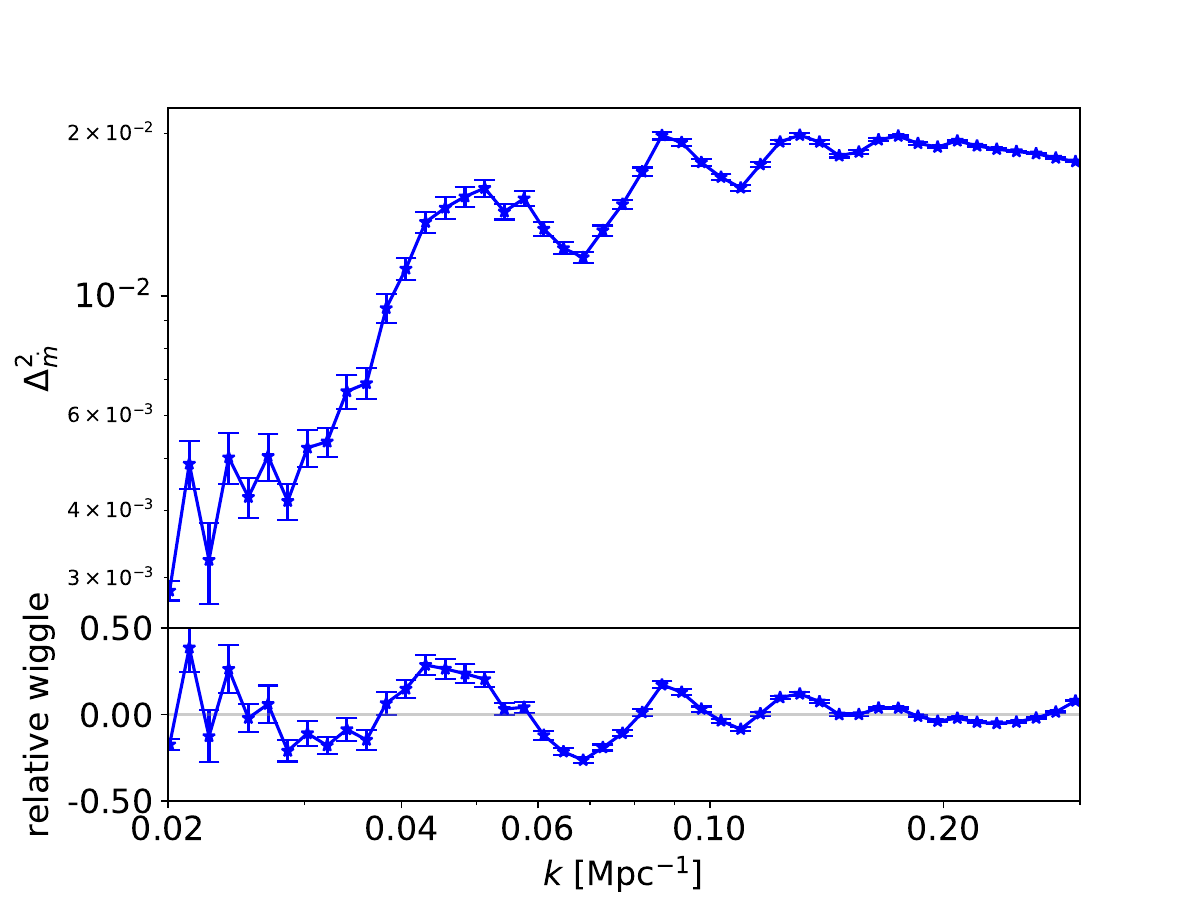}}
\caption{$Top$: A slice of the dimensionless PBH accretion rate field at $z=20$, for $M_{\rm PBH,rec}=200~M_{\odot}$. Here we set $T_{\rm K}=5$ K and $x_{\rm e}=10^{-4}$. 
$Bottom$: The power spectrum of the accretion field, $\Delta_{\dot{m}}^2(k)$ (top sub-panel), and the relative amplitude of the VAOs wiggles, $[\Delta_{\dot{m}}^2(k)-\Delta_{\dot{m},\rm poly}^2(k)]/\Delta_{\dot{m},\rm poly}^2(k)$ (bottom sub-panel). 
}
\label{fig:m_dot_D_P}
\end{figure}

In Fig. \ref{fig:m_dot_D_P}, we show a 2D spatial distribution map of the PBH accretion rate at redshift $z=20$, assuming a constant  gas temperature of $T_{\rm K}=5$ K and ionization fraction $x_{\rm e}=10^{-4}$, for $M_{\rm PBH,rec}=200~M_{\odot}$.
Compared with Fig. \ref{fig:v_bc}, there are strong anti-correlations between the accretion and the relative streaming velocities, in regions with high $v_{\rm db}$ the $\dot{m}$ is significantly reduced. The variations of $\dot{m}$ span several orders of magnitude, which is rather impossible for accretion rate modulated purely by density fluctuations. We also plot the power spectrum of the accretion rate,  $\Delta^2_{\dot{m}}(k)$, in Fig. \ref{fig:m_dot_D_P}, and the relative amplitudes of the wiggles compared to a 5th degree polynomial smooth reference. The power spectrum of the accretion rate clearly exhibits the wiggles, with relative amplitudes up to $\sim 30\%$.

\begin{figure}
\centering
\subfigure{\includegraphics[width=0.45\textwidth]{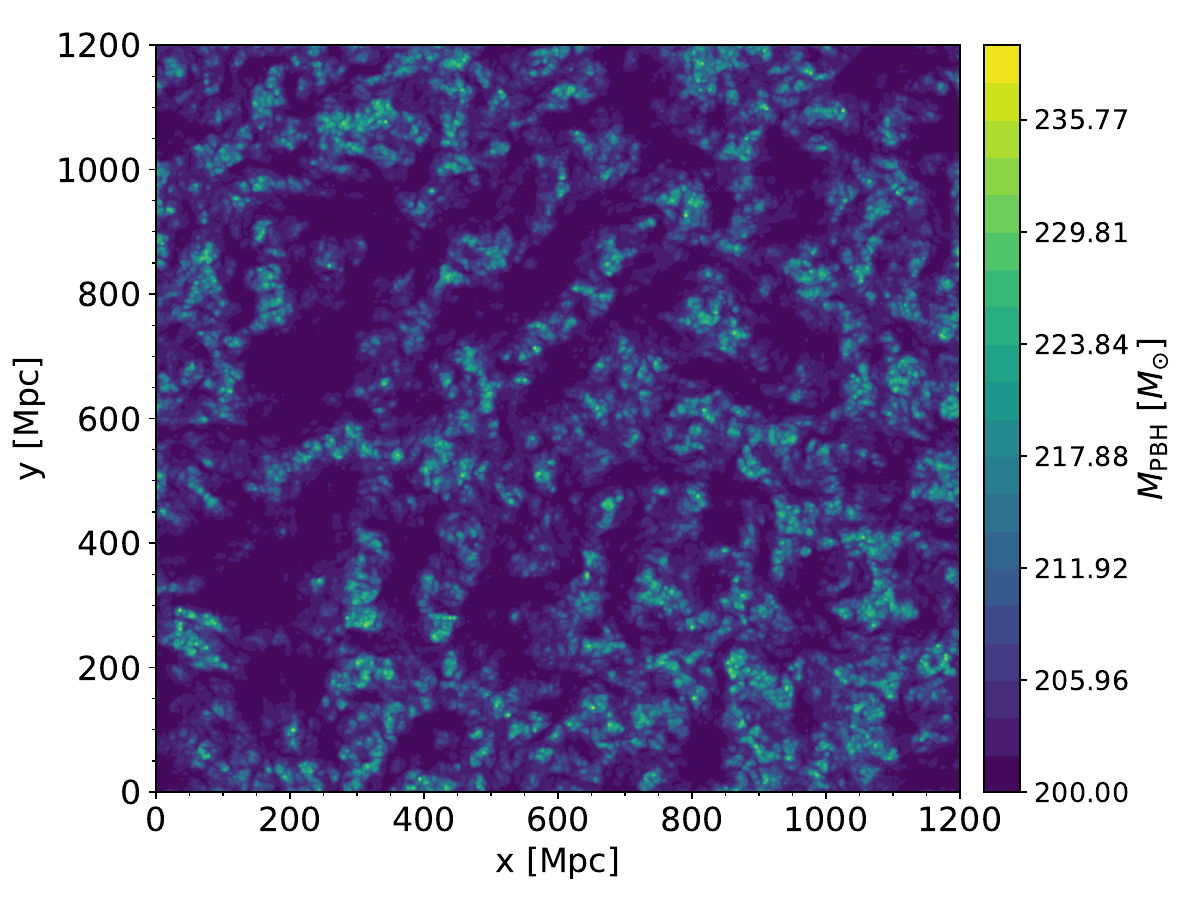}}
\subfigure{\includegraphics[width=0.45\textwidth]{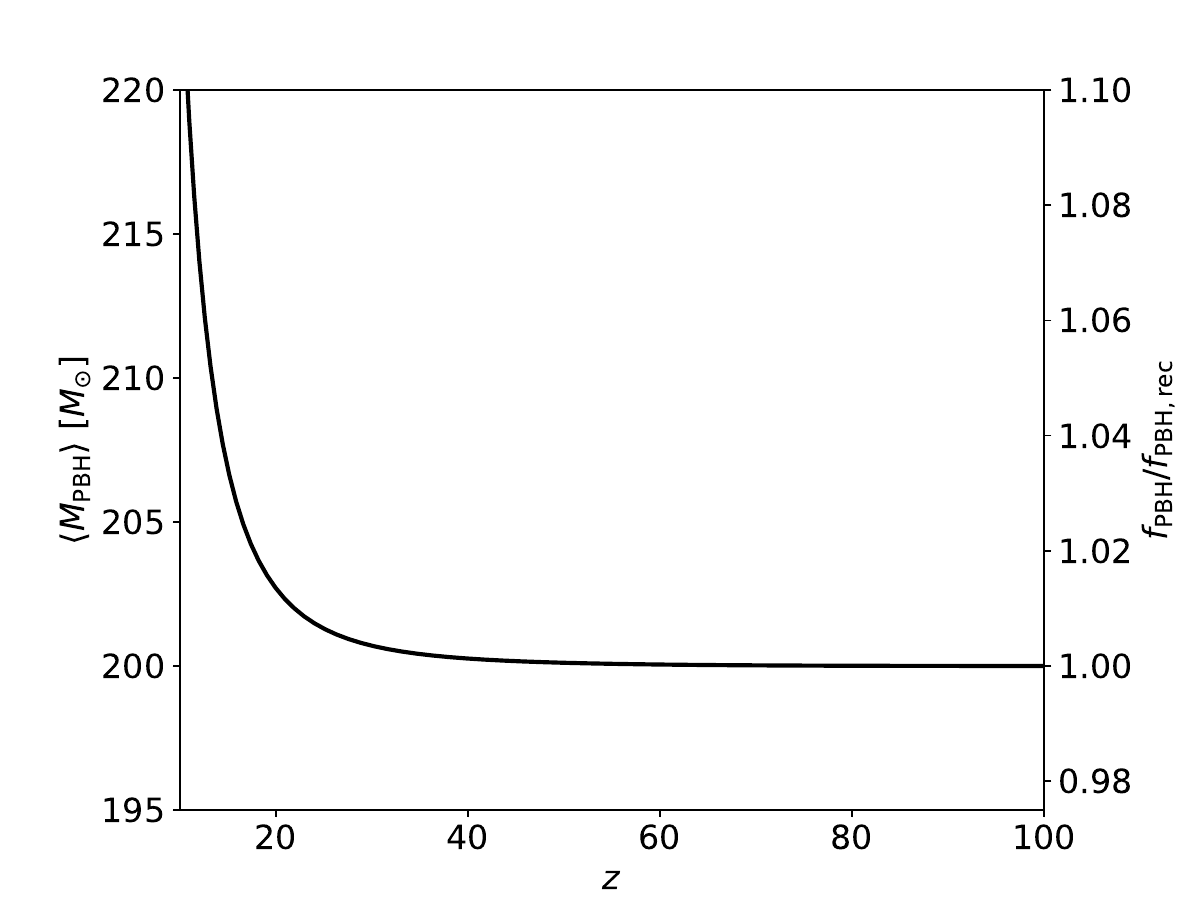}}
\caption{$Top$: A slice of the PBH masses field $M_{\rm PBH}$ at $z=20$. $Bottom$: The growth of the mean PBH mass and fraction in the total DM. Here we set the mass $M_{\rm PBH,rec}=200~M_{\odot}$. 
}
\label{fig:M_PBH_evolution}
\end{figure}

In Fig. \ref{fig:M_PBH_evolution} we show a slice of the $M_{\rm PBH}$ distribution at $z=20$, for $M_{\rm PBH,rec}=200~M_\odot$. Although the initial PBHs are same everywhere, the accretion rate depends on the environment ($v_{\rm db}$ and $\delta_{\rm b}$), so the spatial distribution of $M_{\rm PBH}$ is modulated by the DM-baryon relative streaming velocities as well. However, compared with the $\dot{m}$ map in Fig. \ref{fig:m_dot_D_P}, the fluctuations level of the $M_{\rm PBH}$ map is much smaller. So the fluctuations of the PBHs radiation field must be dominated by accretion rate. In Fig. \ref{fig:M_PBH_evolution} we also show the growth of the mean PBHs mass and fraction in the total DM. Until $z=10$,  this fraction increases by just $\lesssim 10\%$ of the value at the recombination era.

We caution that the sound speed $c_{\rm s}$ depends on the kinetic temperature of the IGM gas, $T_{\rm K}$, while the gas is also heated by the X-ray radiation from PBHs, so $T_{\rm K}$ must be solved self-consistently from the cosmic thermal evolution equations with X-ray heating from PBHs. We will introduce this in next section.

\section{The X-ray radiation from PBHs and the heating to the IGM}\label{sec:X-ray}

For a PBH with accretion rate $\dot{m}$, the bolometric luminosity
\begin{equation}
    L_{\rm bol}=\epsilon_{\rm rad}\dot{m}L_{\rm Edd}.
\end{equation}
Since $L_{\rm bol} \leq L_{\rm Edd}$, we set a hard limit on $\dot{m}$ so that
 $\dot{m}\leq 1/\epsilon_{\rm rad}=10$ always holds. We assume the X-ray luminosity of a PBH with mass $M_{\rm PBH}$ follows the power-law form,
\begin{equation}
L_{\rm X}(\nu_{\rm X},\dot{m},M_{\rm PBH})= A_{\rm X}
 \left(\frac{\nu_{\rm X}}{\nu_{\rm X,\rm min}}\right)^{-\alpha},
\end{equation}
where the normalization factor 
\begin{equation}
A_{\rm X}= \frac{(1-\alpha) \eta_{\rm X} L_{\rm bol}  }{ \nu_{\rm X,\rm min}   \left[\left(\frac{\nu_{\rm 10keV}}{\nu_{\rm X,\rm min}}\right)^{1-\alpha}- \left(\frac{\nu_{\rm 2keV}}{\nu_{\rm X,\rm min}}\right)^{1-\alpha}\right] },
\end{equation}
for which $\nu_{\rm X,min}$ is the minimum frequency of X-ray photons that escape into the IGM, $\nu_{\rm 10keV}$ and $\nu_{\rm 2keV}$ are frequencies of photons with energy 10 keV and 2 keV respectively. We set $\nu_{X,\rm min}=\nu_{\rm 2keV}$. $\eta_{\rm X}$ is the fraction of hard X-ray ($2 - 10$ keV) luminosity in the total bolometric luminosity $L_{\rm bol}$. In observations, the hard X-ray luminosity accounts for roughly $\sim10\%$ of the bolometric luminosity \cite{Shen2020,0708.4308,2004MNRAS.351..169M}, and $\eta_{\rm X}$ could be positively correlated with $\dot{m}$ \cite{Merloni2003,Lusso2012MNRAS}. However for simplicity we take $\eta_{\rm X}=0.1$ in this paper. 

Ignoring the contribution from UV ionizing photons, the time evolution of Hydrogen ionization fraction $x_{\rm HII}$, Helium ionization fraction $x_{\rm HeII}$, and temperature of the IGM, $T_{\rm K}$, follows:
\begin{align}
\frac{{\rm d} x_{\rm HII}}{{\rm d}t} &= I_{\rm H} +  \frac{1}{E_{\rm H}}  \chi_{\rm ion,H}\times  \nonumber \\
&\int_{\nu_{X,\rm min}}^{\rm \infty}4\pi J_{\rm X}(\nu_{\rm X},\vec{r},z)\sigma_{\rm eff}(\nu_{\rm X})d\nu_{\rm X}   \nonumber  \\
\frac{{\rm d} x_{\rm HeII}}{{\rm d}t} &= I_{\rm He} +  \frac{1}{E_{\rm He}}  \chi_{\rm ion,He} \times \nonumber \\
&\int_{\nu_{X,\rm min}}^{\rm \infty}4\pi J_{\rm X}(\nu_{\rm X},\vec{r},z)\sigma_{\rm eff}(\nu_{\rm X})d\nu_{\rm X} \nonumber  \\ 
\frac{{\rm d} T_{\rm K} }{{\rm d}t} &= H_{\rm T}+ \frac{2}{3k_{\rm B}}\chi_{\rm heat}\times \nonumber \\
&\int_{\nu_{\rm X,\rm min}}^{\rm \infty} 4\pi J_{\rm X}(\nu_{\rm X},\vec{r},z)\sigma_{\rm eff}(\nu_{\rm X})d\nu_{\rm X}, \label{eq:heat_ionize}
\end{align}
where
\begin{equation}
\sigma_{\rm eff}=[(1-f_{\rm He})\sigma_{\rm HI}(1-x_{\rm HII})+f_{\rm He}\sigma_{\rm HeI}(1-x_{\rm HeII})].
\end{equation}
$\sigma_{\rm H}$ and $\sigma_{\rm He}$ are ionization cross-sections for Hydrogen and Helium respectively \cite{2003MNRAS.345..379M},
$I_{\rm H}$ and $I_{\rm He}$ are net ionization rates for Hydrogen and Helium, and $H_{\rm T}$ is the net heating rate, contributed by processes in addition to PBHs. Their expressions are given in Eq. (28)  of Ref. \citep{2303.06616}. $\chi_{\rm ion,H}$, $\chi_{\rm ion,He}$ and $\chi_{\rm heat}$ are fractions of X-ray energy deposited into Hydrogen ionization, Helium ionization and heating respectively \cite{VF08}. $E_{\rm H}=13.6~\rm eV$ and $E_{\rm He}=24.6~\rm eV$ are ionization energy for Hydrogen and Helium. $k_{\rm B}$ is the Boltzmann constant and $f_{\rm He}=0.07$ is Helium number fraction in Universe.   
Since $\dot{m}$ depends on $T_{\rm K}$ and $x_{\rm e}$, we solve the thermal evolution of the IGM and $\dot{m}$ self-consistently.

The X-ray radiation intensity at position $\vec{r}$ and redshift $z$
\begin{align}
    J_{\rm X}(\nu_{\rm X},\vec{r},z)&= \frac{(1+z)^3}{4\pi} \int_{0}^{\infty} 
\tilde{\epsilon}'(\nu'_X,\vec{r},R'_{\rm c}) e^{-\tilde{\tau}'(\nu_{\rm X},\vec{r},R'_c,z)}\frac{dR'_{\rm c}}{1+z'}, 
\end{align}
where $R'_{\rm c}$ is the comoving distance from redshift $z'$ to redshift $z$, $\nu'_X=\frac{(1+z')}{(1+z)}\nu_{\rm X}$ is the emitted photon frequency. The mean emissivity in a shell at radius $R'_{\rm c}$ to $\vec{r}$, at $z'$,
\begin{align}
&\tilde{\epsilon}'(\nu'_X,R'_c,\vec{r})=n_{\rm PBH} \times \nonumber \\
&\frac{1}{4\pi}\int_{4\pi} L_{\rm X}(\nu'_X,\dot{m}(R'_{\rm c},\vec{\theta}),M_{\rm PBH}(R'_{\rm c},\vec{\theta}))d\Omega(\vec{\theta}),
\end{align}
where $n_{\rm PBH}=\rho_{\rm c}\Omega_{\rm DM}f_{\rm PBH,rec}/M_{\rm PBH,rec}$ is the comoving number density of PBHs, $\Omega_{\rm DM}$ denotes the total DM density,
and $M_{\rm PBH,rec}(R'_{\rm c},\vec{\theta})$ is the mass of PBHs located at the direction $\vec{\theta}$  and distance $R'_{\rm c}$ from the center $\vec{r}$. $\tilde{\tau}'(\nu_{\rm X}, \vec{r},R'_{\rm c},z)$ is the mean optical depth for photons with emitted frequency $\nu'$ at redshift $z'$, travel a comoving distance $R'_{\rm c}$  and arrive at $\vec{r}$ at redshift $z$.  It is
 \begin{equation}
 \begin{aligned}
     &\tilde{\tau}'(\nu_{\rm X},\vec{r}, R'_{\rm c},z )\\
     &\approx\int_{0}^{R'_{\rm c}}\overline{n}_{\rm b}(1-f_{\rm He})(1+\tilde{\delta}''_{\rm b})(1+z'')^{\rm 3}\sigma_{\rm H}(\nu''_X)\frac{dR_{\rm c}''}{1+z''}  \\
     &+\int_{0}^{R'_{\rm c}}\overline{n}_{\rm b} f_{\rm He}(1+\tilde{\delta}''_{\rm b})(1+z'')^{\rm 3}\sigma_{\rm He}(\nu''_X)\frac{dR_{\rm c}''}{1+z''}
\end{aligned}
 \end{equation}
where $R_{\rm c}''$ is the comoving distance from $z''$ to $z$, 
$\overline{n}_{\rm b}$ is the cosmic comoving baryon number density, 
$\tilde{\delta}''_{\rm b}(z'',R''_{\rm c},\vec{r})$ is mean density contrast on a spherical shell of radius $R_{\rm c}''$ centered at $\vec{r}$ at redshift $z''$.   
 
\begin{figure}
\centering
\subfigure{\includegraphics[width=0.45\textwidth]{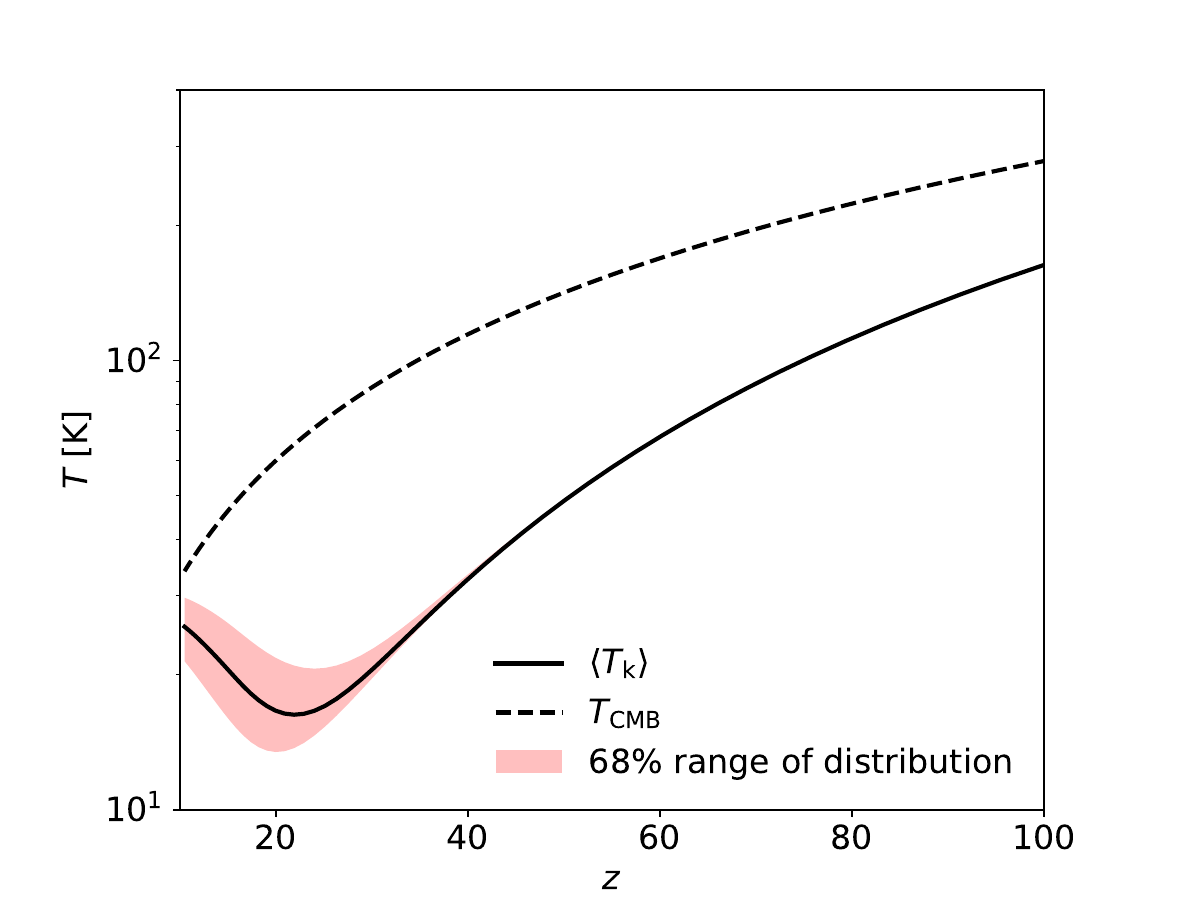}}
\caption{The evolution of the mean and scattering of the IGM temperature under the influence of PBHs radiation, assuming $M_{\rm PBH,rec}=200~M_{\odot}$ and $f_{\rm PBH,rec}=10^{-8}$. The black dashed line is the CMB temperature.
} 
\label{fig:Tk_and_xe}
\end{figure}

Both the $\tilde{\epsilon}'$ and the $\tilde{\delta}''_{\rm b}$ are calculated by smoothing the fields in Fourier space \citep{2011MNRAS.411..955M}, so the computations of $J_{\rm X}$ and $\tilde{\tau}'$ can be efficient.
Fig. \ref{fig:Tk_and_xe} shows the evolution of mean gas temperature with the PBHs mass $M_{\rm PBH,rec}=200~M_{\odot}$ and PBHs fraction $f_{\rm PBH,rec}=10^{-8}$ at the recombination era, compared with the CMB temperature.
 Before $z\sim 20$ the heating is negligible, since then the IGM temperature starts to increase. PBHs heating results in an increasingly dispersed temperature distribution.

\begin{figure}
\centering
\subfigure{\includegraphics[width=0.45\textwidth]{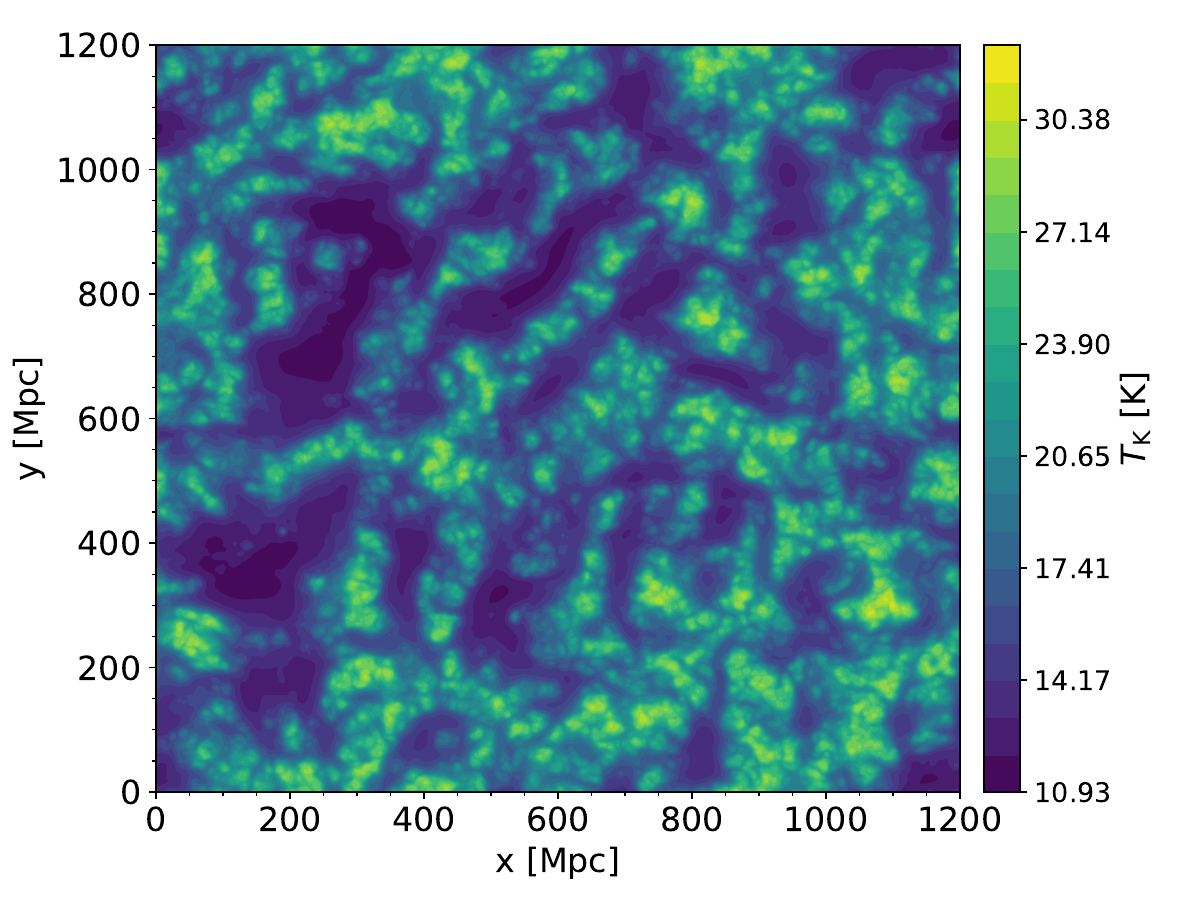}}
\subfigure{\includegraphics[width=0.45\textwidth]{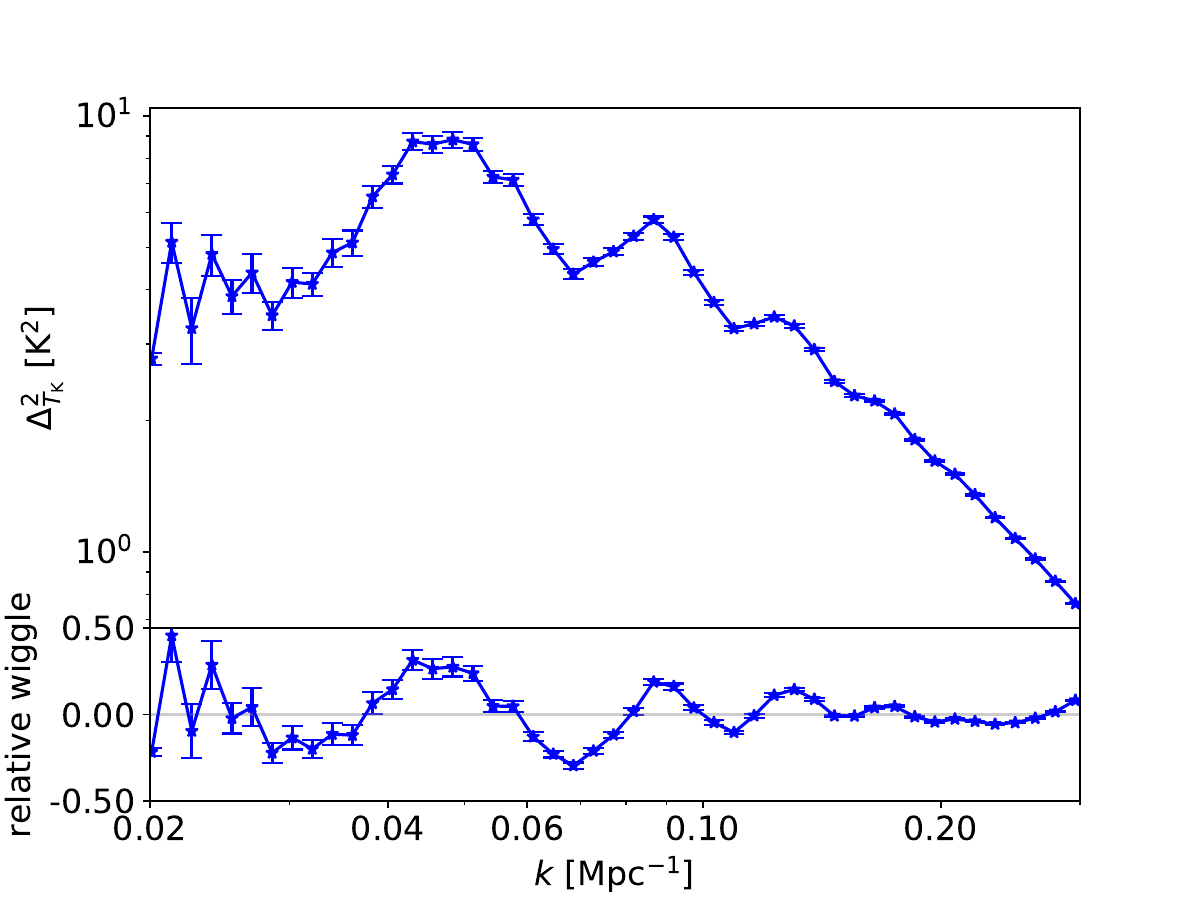}}
\caption{$Top$: A slice of the gas temperature field at $z=20$. The IGM is heated by PBHs with $M_{\rm PBH,rec}=200~M_{\odot}$ and $f_{\rm PBH,rec}=10^{-8}$. $Bottom$: The power spectrum of the gas temperature field, $\Delta_{T_{\rm K}}^2(k)$ (top sub-panel), and the relative amplitude of the VAOs wiggles (bottom sub-panel). 
} 
\label{fig:Tk_D_P}
\end{figure}

Because of the optical depth, X-ray photons would continuously attenuate in  their propagation. Therefore, the X-ray radiation produced by PBHs could only heat and ionize the IGM around PBHs. This induces VAOs features in the power spectrum of the IGM temperature field.
In  Fig. \ref{fig:Tk_D_P}, we show a slice of the IGM temperature field and the power spectrum at $z=20$, for $M_{\rm PBH,rec}=200~M_{\odot}$ and $f_{\rm PBH,rec}=10^{-8}$. 
We can see obvious large-scale inhomogeneity in the temperature field, and VAOs wiggles in the power spectrum. 
Interestingly, compared with the power spectrum of accretion rate $\dot{m}$ in Fig. \ref{fig:m_dot_D_P}, the IGM temperature power spectrum damps gradually at small scales.
This damping arises because X-rays have a finite propagation length, fluctuations on scales below the their mean free path, $\lambda_{\rm X}$, are damped. The X-ray photons number spectrum weighted $physical$ mean free path can be estimated as 
\begin{align}
    \lambda_{\rm X}&\sim \frac{(1+z)^{-3}\int \frac{1}{h_{\rm P}\nu_{\rm X}}L_{\rm X}(\nu_{\rm X}, \dot{m}, M_{\rm PBH}) d \nu_{\rm X} }{\int  \overline{n}_{\rm b} \sigma_{\rm eff}(\nu_{\rm X}) \frac{1}{h_{\rm P}\nu_{\rm X}} L_{\rm X}(\nu_{\rm X},\dot{m},M_{\rm PBH})d\nu_{\rm X}}, 
\end{align}
where $h_{\rm p}$ is the Planck constant. 
At $z=20$ $\lambda_{\rm X}\simeq6$ Mpc.
This $physical$ length is smaller than the Hubble scale at this redshift, implying that the emitted X-ray photons can be efficiently absorbed within a cosmological timescale at that redshift, leading to heating effect is concentrated around the radiation sources. The corresponding comoving wavenumber is $k_{\rm mfp}\sim0.05~{\rm Mpc}^{-1}$. 
Modes with $k\gtrsim k_{\rm mfp}$ are progressively damped. Therefore, the overall amplitude of the temperature power spectrum declines toward higher $k$.
However, we find that the relative amplitude of the wiggles remains essentially unchanged, see the bottom panel of Fig. \ref{fig:Tk_D_P}.

\section{The 21 cm signal}\label{sec:21cm}

\subsection{The VAOs features in the  21 cm power spectrum in Dark Ages}
\label{subsction:VAO_21cm}

The 21 cm brightness temperature ~\cite{21cm_review2_2012RPPh...75h6901P}:
\begin{eqnarray}
\delta T_{\rm b} &=& 27 x_{\rm HI}(1+\delta_{\rm b})\left(\frac{\Omega_{\rm b} h^2}{0.023}\right)\left( \frac{0.15}{\Omega_{\rm m} h^2} \frac{1+z}{10} \right)^{1/2} \nonumber \\
& \times & \left( \frac{T_{\rm S}-T_{\rm CMB}}{T_{\rm S}}\right)~[\rm mK],
\label{eq:deltaTb}
\end{eqnarray}    
where $T_{\rm CMB}$ is the CMB temperature at $z$. 
The spin temperature  
\begin{equation}
T_{\rm S}^{-1}=\frac{T_{\rm CMB}^{-1} +x_\alpha T_{\rm K}^{-1} +x_{\rm c}T_{\rm K}^{-1}}{1+x_\alpha+x_{\rm c}},    
\end{equation}
where $x_\alpha$ and $x_{\rm c}$ are coupling coefficients due to Ly$\alpha$ scattering and collision respectively \cite{2011MNRAS.411..955M}. 
In Dark Ages, we ignore the astrophysical objects like stars, galaxies and black holes, the only sources (ignore the Ly$\alpha$ radiation directly from the accreting region) of the Ly$\alpha$ radiation is the deposition of X-ray radiation from PBHs. The corresponding Ly$\alpha$ background flux is 
\begin{align}
    J_{\alpha}(\vec{r},z) &=
    \frac{1}{h_{\rm p}\nu_{\alpha}}\frac{(1+z)^3}{4\pi}\frac{c}{H(z)}\frac{\chi_{\alpha}}{\nu_{\alpha}} \nonumber\\
    &\quad \times
    \int_{\nu_{X,\rm min}}^{\infty}
    4\pi J_{\rm X}(\nu_{\rm X},\vec{r},z)\overline{n}_{\rm b}(1+\delta_{\rm b})\sigma_{\rm eff}(\nu_{\rm X}) \,d\nu_{\rm X},
\end{align}
where $H(z)$ is the Hubble parameter, $\nu_{\alpha}$ is the frequency of Ly$\alpha$ photons and $\chi_{\alpha}$ is fraction of X-ray energy deposited into Ly$\alpha$ photons \cite{VF08}.
Moreover, the X-ray also heats the IGM. Therefore the 21 cm signal is modulated by the distribution of $\dot{m}$, and the $v_{\rm db}$.

\begin{figure}
\centering
\subfigure{\includegraphics[width=0.45\textwidth]{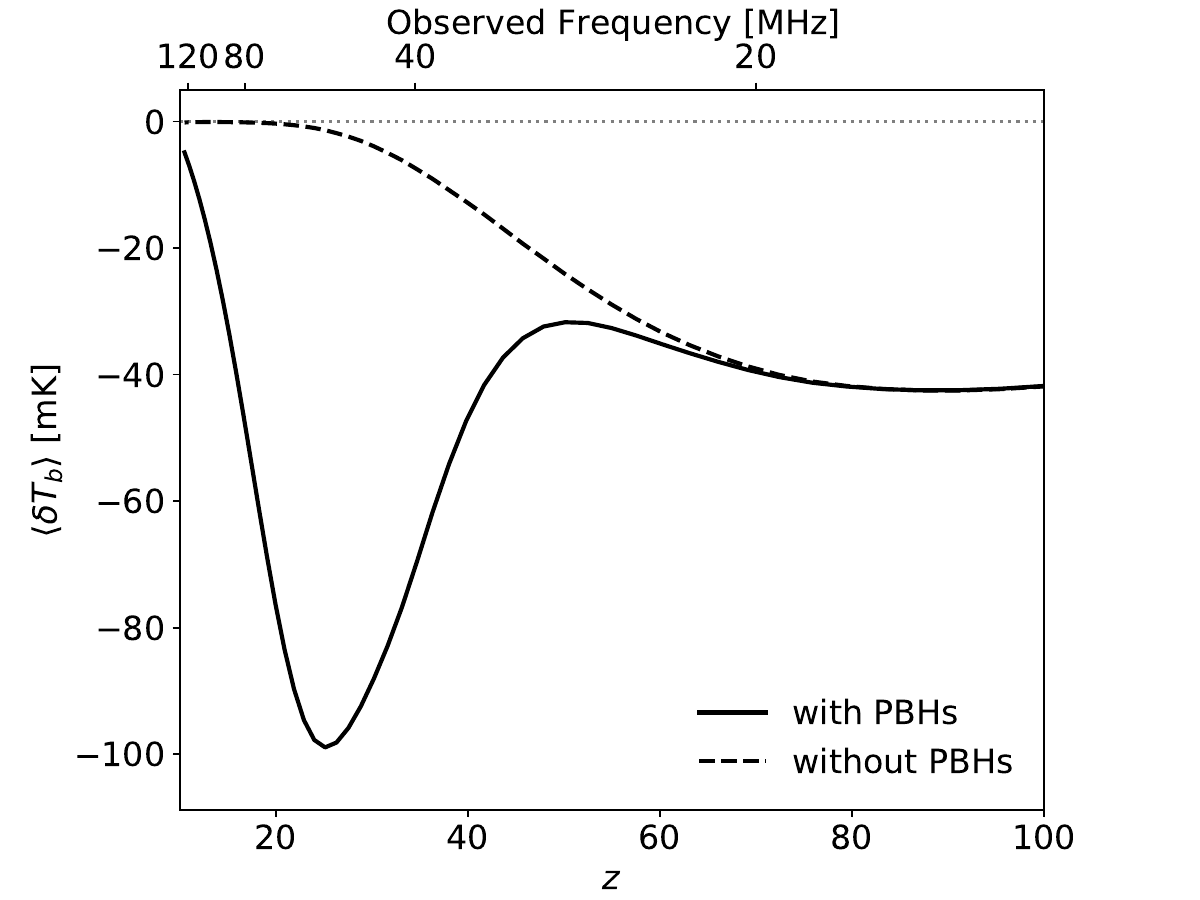}}
\caption{The redshift evolution of the mean 21 cm brightness temperature in the absence/presence of PBHs with $M_{\rm PBH,rec}=200~M_\odot$ and $f_{\rm PBH,rec}=10^{-8}$.
} 
\label{fig:T21_noCRB}
\end{figure}

Fig. \ref{fig:T21_noCRB} shows the evolution of the global 21 cm brightness temperature in the absence/presence of PBHs with $M_{\rm PBH,rec}=200~M_{\odot}$ and $f_{\rm PBH,rec}=10^{-8}$. In the presence of PBHs, the 21 cm absorption signal becomes deeper at $z\lesssim 50$, thanks to the PBHs-induced Ly$\alpha$ scattering. At $z\lesssim 25$, the 21 cm absorption becomes weaker because of the PBHs-induced X-ray heating. So Ly$\alpha$ scattering takes effect earlier than X-ray heating.

\begin{figure*}
\centering
\subfigure{\includegraphics[width=0.45\textwidth]{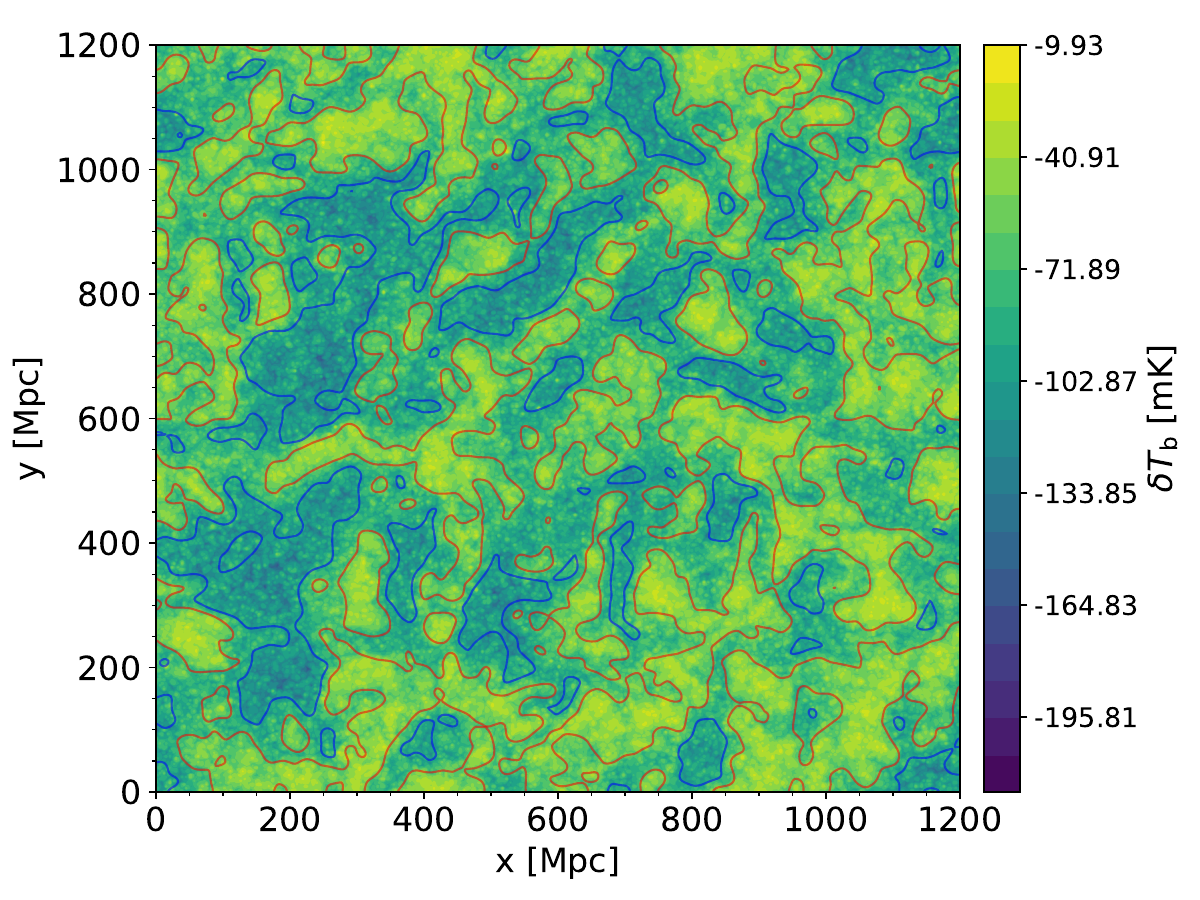}}
\subfigure{\includegraphics[width=0.45\textwidth]{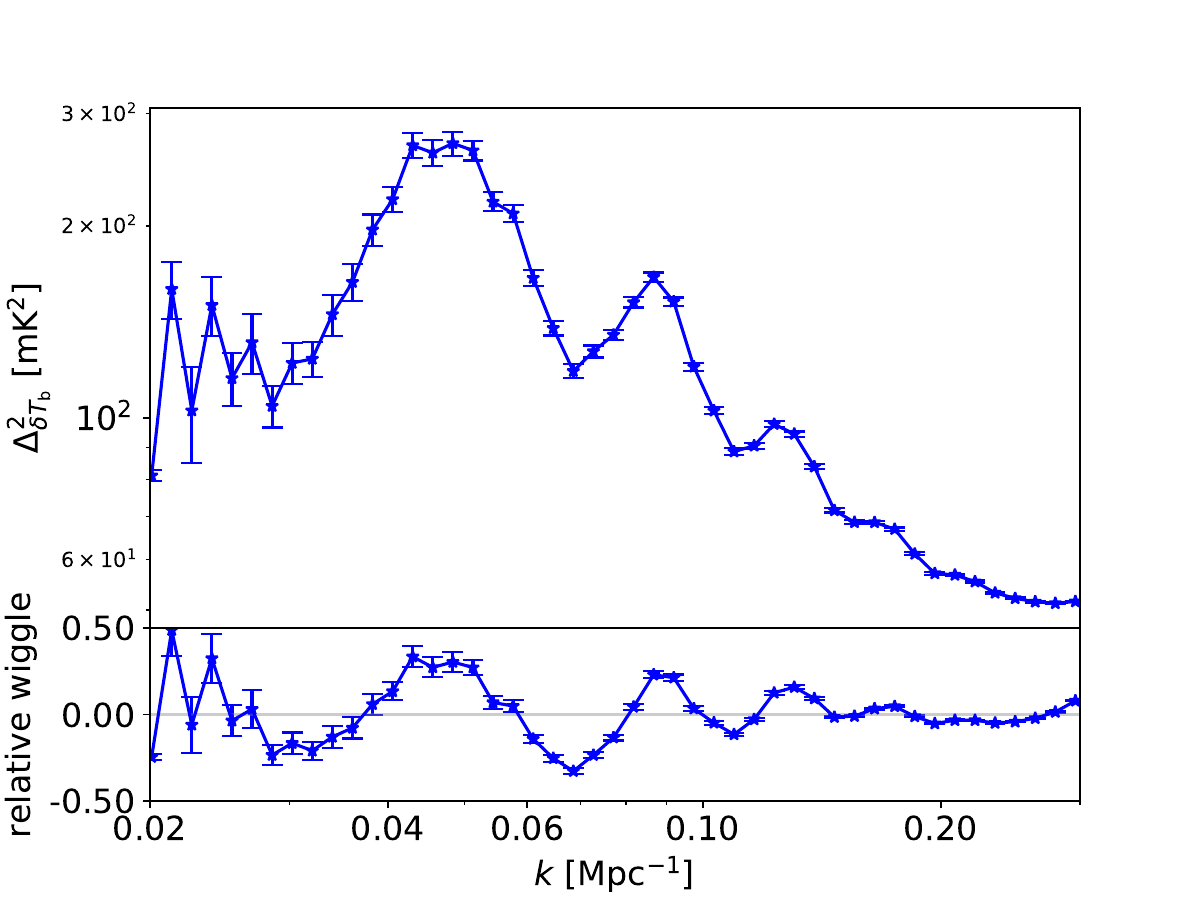}}
\subfigure{\includegraphics[width=0.45\textwidth]{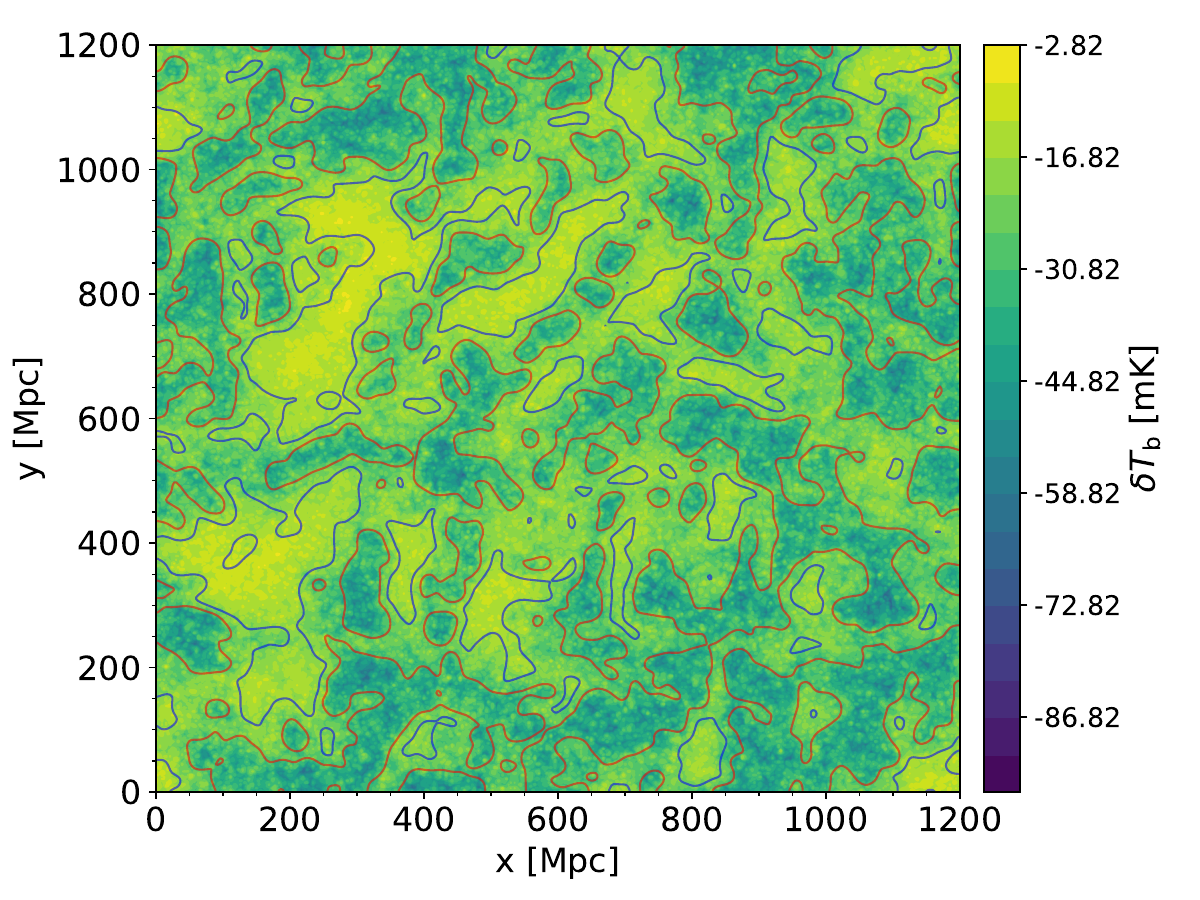}}
\subfigure{\includegraphics[width=0.45\textwidth]{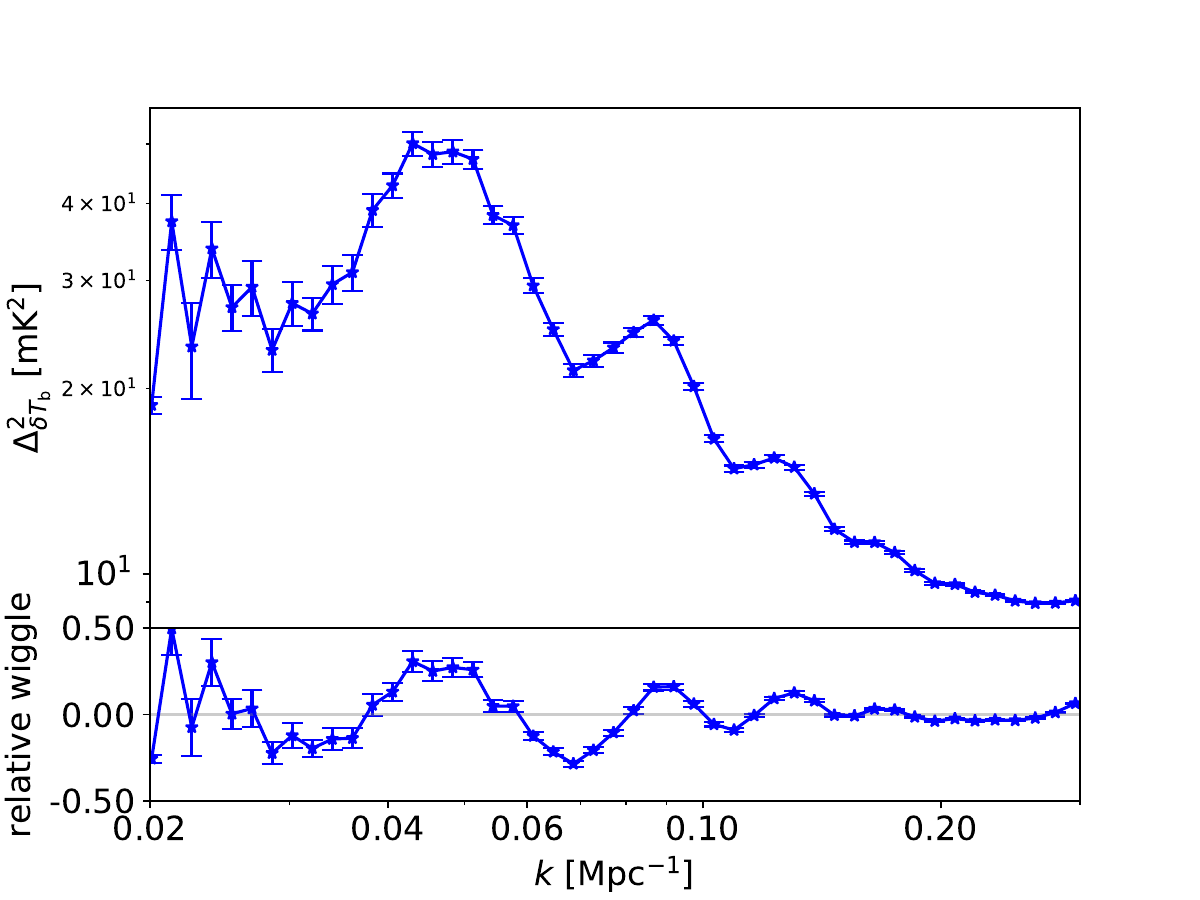}}
\caption{$Left$: Slices of the 21 cm brightness temperature field at $z=20$ (corresponding to observed frequency $\nu_{\rm obs}=67~\rm MHz$), for PBHs with $M_{\rm PBH,rec}=200~M_{\odot}$. $Right$: The corresponding power spectra of the 21 cm signal, $\Delta_{\rm 21}^{\rm 2}(k)$ (top sub-panels) and the relative amplitude of the VAOs wiggles 
$[\Delta_{21}^2(k)-\Delta_{21,\rm poly}^2(k)]/\Delta_{21,\rm poly}^2(k)$ (bottom sub-panel).  The $Top$ and $Bottom$ panels correspond to $f_{\rm PBH,rec}=10^{-8}$ and $f_{\rm PBH,rec}=10^{-10}$, respectively. 
In the top left panel, regions with $\dot{m}\ge 0.08$ are marked by red contours, regions with $\dot{m}\le 0.04$ by blues. In bottom left panel, regions with $\dot{m} \ge 0.15$ are marked by red contours, regions with $\dot{m}\le 0.05$ by blues. Note that we do not use same $\dot{m}$ criteria for contour lines in the top and bottom panels, because for  the lower PBHs fraction  case, the feedback is weaker so the accretion rate is systematically higher.}
\label{fig:T21_noCRB_D_P}
\end{figure*}

In Fig. \ref{fig:T21_noCRB_D_P} we show the 21 cm signal maps and the corresponding power spectra at $z=20$ for a higher PBH fraction in the total DM case ($f_{\rm PBH,rec}=10^{-8}$) and a lower fraction case ($f_{\rm PBH,rec}=10^{-10}$) respectively, assuming $M_{\rm PBH,rec}=200~M_{\odot}$. In the power spectrum panels, we see there are strong VAOs wiggles in the 21 cm power spectrum, with relative amplitude up to $\sim 30\%$. Generally, the 21 cm power spectrum exhibits a rising slope toward smaller scales (larger $k$), tracing the matter distribution. However, in our case, the 21 cm signal is mainly modulated by the spatial distribution of  X-ray, which is determined by $\dot{m}$ and $v_{\rm db}$, while the PBHs themselves are assumed to have uniform distribution. As a result, the power spectrum decreasing toward to smaller scales.

Regarding the fluctuations of the 21 cm signal, both the Ly$\alpha$ scattering and the X-ray heating can change the spin temperature; however, for same amount of X-rays, the Ly$\alpha$ scattering works more efficiently. As a result, when $f_{\rm PBH,rec}$ is smaller, at $z=20$ the IGM is not yet obviously heated, the 21 cm fluctuations mainly reflect the inhomogeneity of the energies deposited into Ly$\alpha$ photons. However, for larger $f_{\rm PBH,rec}$, at $z=20$ the Ly$\alpha$ scattering is almost saturated, 21 cm signal fluctuations mainly reflect the inhomogeneity of the energies deposited into IGM. This can be demonstrated by comparing the 21 cm maps for these two cases.

For the higher PBHs fraction case, in regions with higher accretion rates, Ly$\alpha$ scattering has already saturated before $z=20$, and X-ray heating has intense impacts on the IGM temperature. In such regions the 21 cm absorption is shallower. In regions with lower accretion rates, however, although the Ly$\alpha$ scattering is also saturated or at least very efficient, X-ray heating is almost negligible. Hence in such regions the 21 cm absorption is deeper. For the lower PBHs fraction case, in regions with higher accretion rates, at $z=20$ the Ly$\alpha$ scattering works efficiently but X-ray heating is negligible, so the 21 cm absorption is deeper. This is like the regions with lower accretion rates in the case of higher PBHs fraction. In regions with lower accretion rates, both Ly$\alpha$ scattering and X-ray heating are less efficient, as a result the 21 cm absorption is shallower. In summary, for $M_{\rm PBH,rec}=200~M_\odot$ and at $z=20$, for the higher PBHs fraction case, the 21 cm absorption is anti-correlated with accretion rate, while for the lower fraction case the 21 cm absorption is positively correlated with the accretion rate. To highlight this point, in Fig. \ref{fig:T21_noCRB_D_P} we mark the regions with higher accretion rates and lower accretion rates in the 21 cm signal fields.
Interestingly, although the spatial fluctuations in these two cases exhibit opposite trends, the corresponding power spectra have similar shapes and show very similar VAOs features.

From the above, we can imagine that there must exist an intermediate PBHs fraction between these two cases, for which the 21 cm absorption is almost independent of PBHs accretion rate and the VAOs features disappear. We will analyze this in following paragraphs.

\begin{figure}
\centering
\subfigure{\includegraphics[width=0.45\textwidth]{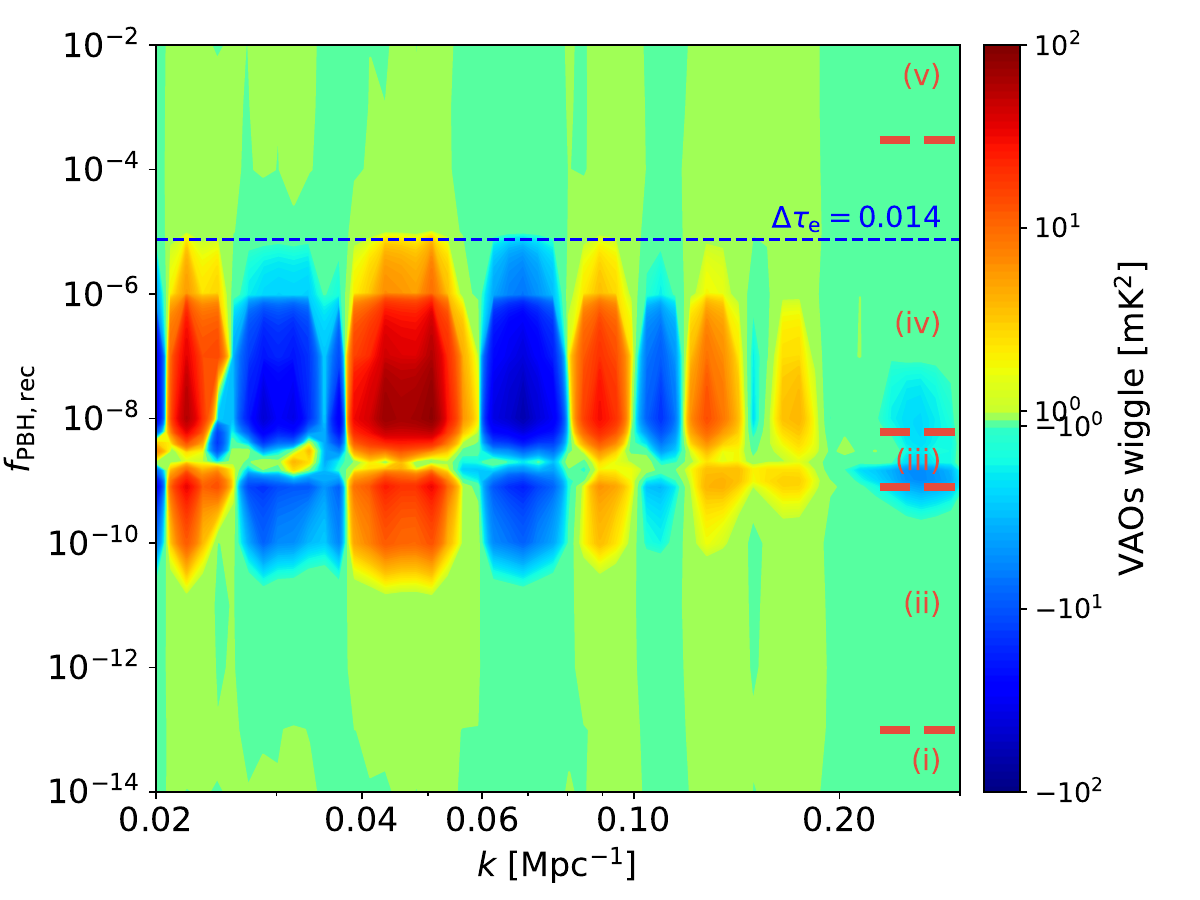}}
\subfigure{\includegraphics[width=0.45\textwidth]{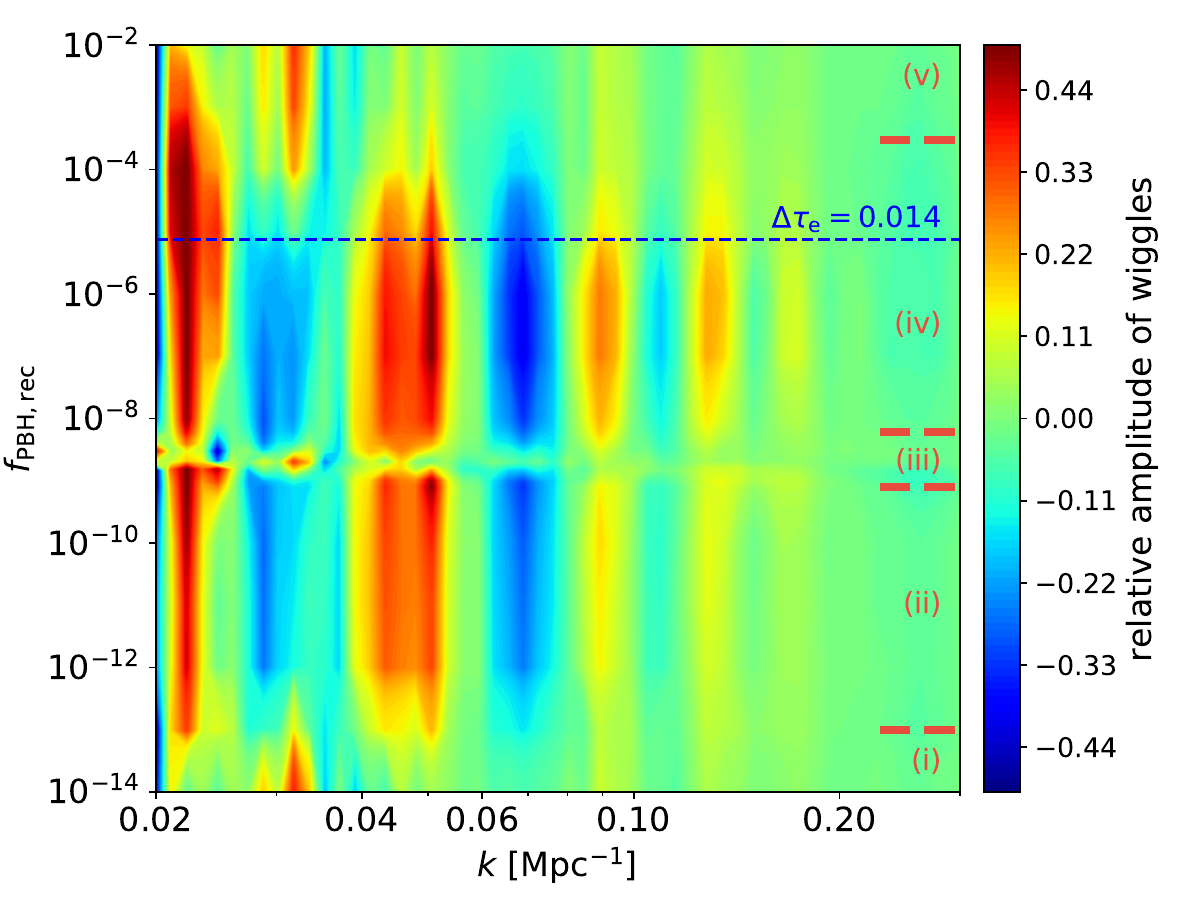}}
\subfigure{\includegraphics[width=0.45\textwidth]{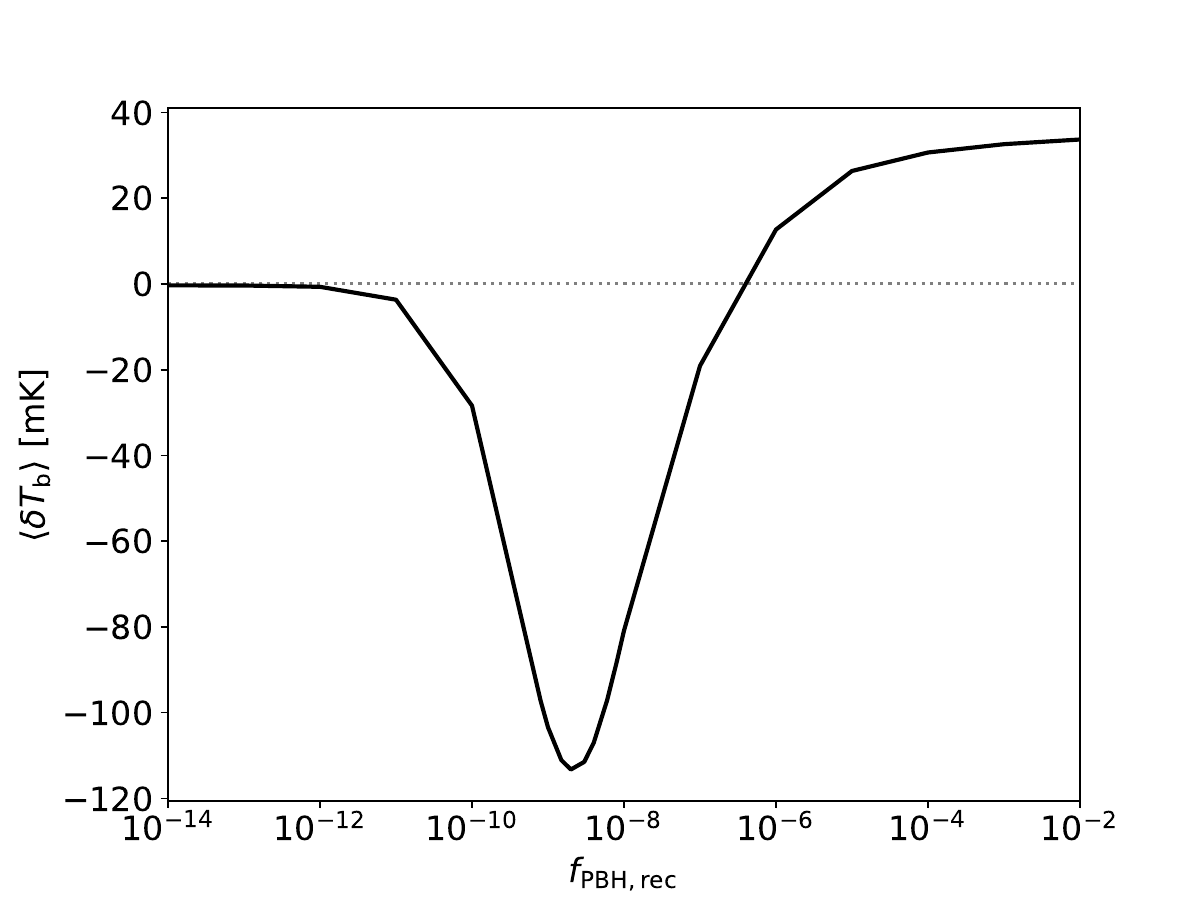}}
\caption{ 
{\it Top:} The VAOs wiggles in the 21 cm power spectrum, $\Delta_{21}^2(k)-\Delta_{21,\rm poly}^2(k)$, at $z=20$, as a function of  $f_{\rm PBH,rec}$, assuming $M_{\rm PBH,rec}=200~M_{\odot}$. The blue dashed line refers to the critical PBH fraction above which the PBHs-induced CMB scattering optical depth breaks the 2$\sigma$ uncertainties in {\tt Planck18} measurements.
{\it Middle:} The relative amplitude of the above wiggles, $[\Delta_{21}^2(k)-\Delta_{21,\rm poly}^2(k)]/\Delta_{21,\rm poly}^2(k)$.  
{\it Bottom:} The mean 21 cm brightness temperature at $z=20$, as a function of $f_{\rm PBH,rec}$, for $M_{\rm PBH,rec}=200~M_\odot$.
}
\label{fig:contourf_Pk_f_PBH}
\end{figure}

In the top and middle panels of Fig. \ref{fig:contourf_Pk_f_PBH}, we show the VAOs wiggles and their relative amplitude at $z=20$, as a function of $f_{\rm PBH,rec}$, for $M_{\rm PBH,rec}=200~M_\odot$. The VAOs wiggles in top panel are weaker when $f_{\rm PBH,rec}\lesssim 10^{-10}$, because of the weaker Ly$\alpha$ scattering and X-ray heating; and when $f_{\rm PBH,rec}\gtrsim 10^{-6}$ as the strong X-ray heating release the 21 cm signal from spin temperature. However, for relative amplitude in the middle panel, the wiggles are visible for $10^{-13}\lesssim f_{\rm PBH,rec}\lesssim 10^{-3} $. As the PBHs fraction increases, the impact of PBHs becomes progressively stronger. The results at $z=20$ are classified into five characteristic phases:

\begin{enumerate}[(i)]

\item For $f_{\rm PBH,rec}\lesssim 10^{-13}$, the impacts of PBHs are negligible. $T_{\rm K}$ always follows the adiabatic-cooling track and $T_{\rm S}\sim T_{\rm CMB}$ when the collisional coupling becomes inefficient at $z\lesssim 30$. In this case there are no VAOs features in the 21 cm signal.

\item For $10^{-13}\lesssim f_{\rm PBH,rec}\lesssim 2\times 10^{-9}$, the Ly$\alpha$ scattering effect starts to become effective, but the X-ray heating effect is still inefficient. As a result the anisotropy of the 21 cm signal is modulated by both the $\delta_{\rm b}$ and the $x_\alpha$ that is tightly related to $\dot{m}$ and $v_{\rm db}$. Therefore the VAOs features appear in the 21 cm signal.
Moreover, since the IGM is not yet heated, the 21 cm signal can reach the maximum absorption $\sim -200$ mK.

\item For $f_{\rm PBH,rec}\sim 2\times 10^{-9}$, the Ly$\alpha$ scattering becomes saturated, however the IGM heating is still negligible, therefore $T_{\rm S}\sim T_{\rm K}\sim T_{\rm K}^{\rm ad}$, where $T_{\rm K}^{\rm ad}$ is the IGM temperature in the absence of radiative heating and cooling. The spin temperature is independent of PBHs accretion rate,
and the VAOs features disappear completely. This only occurs within a very narrow range of $f_{\rm PBH,rec}$.

\item For $2\times 10^{-9} \lesssim f_{\rm PBH,rec} \lesssim 10^{-3}$, the X-ray heating starts to become effect, since the Ly$\alpha$ scattering is already saturated, $T_{\rm S}\sim T_{\rm K}$, and $T_{\rm K}$ is modulated by $\dot{m}$ and $v_{\rm db}$. The anisotropy of the 21 cm signal purely replies on the anisotropy of $T_{\rm K}$. VAOs features appear again.

\item For $f_{\rm PBH,rec} \gtrsim 10^{-3}$, X-ray heating is much effcient so that $T_{\rm S}\sim T_{\rm K} \gg T_{\rm CMB}$, as a result the 21 cm signal is independent of the spin temperature and almost traces the $\delta_{\rm b}$. VAOs features disappear again.

\end{enumerate}

In above, the transition $f_{\rm PBH,rec}$ values between different phases are for $M_{\rm PBH,rec}=200~M_\odot$. For smaller $M_{\rm PBH,rec}$, the transition $f_{\rm PBH,rec}$ is larger, while for larger $M_{\rm PBH,rec}$, the transition $f_{\rm PBH,rec}$ is smaller.
Note that for $M_{\rm PBH,rec}=200~M_\odot$, VAOs features appear for PBHs fraction  as low as $\sim 10^{-13}$, 
highlighting the remarkable sensitivity of the 21 cm signal to even a tiny PBHs population. This is well below the current observational upper limit constraints on  $f_{\rm PBH}$ for PBHs with comparable mass \cite{2020PhRvR...2b3204S,2017PhRvD..95d3534A,2017PhRvD..96h3524P,2008ApJ...680..829R,Hektor2018PRD}.

In principle, there are two scenarios can explain the disappearance of VAOs features in phase (iii). In the first one, in regions with higher accretion rates, the Ly$\alpha$ scattering is strong and the gas is heated to close to CMB temperatures, so $T_{\rm S}\sim T_{\rm K}\sim T_{\rm CMB}$, the 21 cm signal is weak; while in regions with lower accretion rates, both Ly$\alpha$ scattering and the X-ray heating is weak,  so $T_{\rm S}\sim T_{\rm CMB}$, the 21 cm signal is also weak. In the second one, in regions both with higher accretion rates and  lower accretion rates, the Ly$\alpha$ scattering has already been saturated, but X-ray heating is almost negligible everywhere. So $T_{\rm S}\sim T_{\rm K}\sim T_{\rm K}^{\rm ad} \ll T_{\rm CMB}$, so the fluctuations of the 21 cm signal is independent of the PBHs accretion rate, VAOs features disappear. In the first scenario, the mean 21 cm signal for this phase should be close to zero, in the second scenario, however, the mean 21 cm signal should be in deep absorption. From the bottom panel of  Fig. \ref{fig:contourf_Pk_f_PBH}, we find the second scenario is the case. This highlights the value of joint global signal-power spectrum observations.

The X-ray from PBHs not only heats the IGM, but also ionizes it. This increases the Thomson scattering optical depth to the CMB by, 
\begin{equation}
    \Delta\tau_{\rm e}(>z)=\int_{z}^{\infty}\sigma_{\rm T}\overline{n}_{\rm H}(z')\Delta x_{\rm e}(z')\frac{cdz'}{H(z')(1+z')}
\end{equation}
where $\Delta x_{\rm e}$ is the fraction of extra IGM electrons generated by the X-ray from PBHs, $\overline{n}_{\rm H}$ is the cosmic Hydrogen number density. 
$\Delta \tau_{\rm e}$ puts constraints on $f_{\rm PBH,rec}$ \cite{2021PhRvD.104f3534J}. In Fig. \ref{fig:contourf_Pk_f_PBH}, we mark the $f_{\rm PBH,rec}$ for which  $\Delta\tau_{\rm e}=0.014$, the $2\sigma$ uncertainty level of the {\tt Planck} observations \cite{2020A&A...641A...6P}.

\subsection{The redshift-evolution of the VAOs features}

For given $M_{\rm PBH,rec}$ and $f_{\rm PBH,rec}$, since the accretion rate increases with decreasing redshift, the VAOs features evolve with redshift as well. In idea case, there are also  five stages similar to Sec. \ref{subsction:VAO_21cm}: a) When the redshift is high enough, the accretion rate is still small and the influence on 21 cm signal is negligible, there are no VAOs features. b) Then the Ly$\alpha$ radiation from PBHs starts to change the $T_{\rm S}$ but the X-ray heating is still negligible, VAOs features appear and purely reply on the anisotropy of Ly$\alpha$ radiation. c) Next, Ly$\alpha$ scattering becomes saturated but X-ray heating is still too weak, so the VAOs features disappear. d) When X-ray heating starts to work, the IGM is heated, VAOs features appear again and purely rely on the anisotropy of X-ray heating. e) Finally when the IGM is heated to $T_{\rm K}\gg T_{\rm CMB}$, the 21 cm signal is independent of $T_{\rm S}$ and VAOs features disappear completely.

\begin{figure*}
\centering
\subfigure{\includegraphics[width=0.90\textwidth]{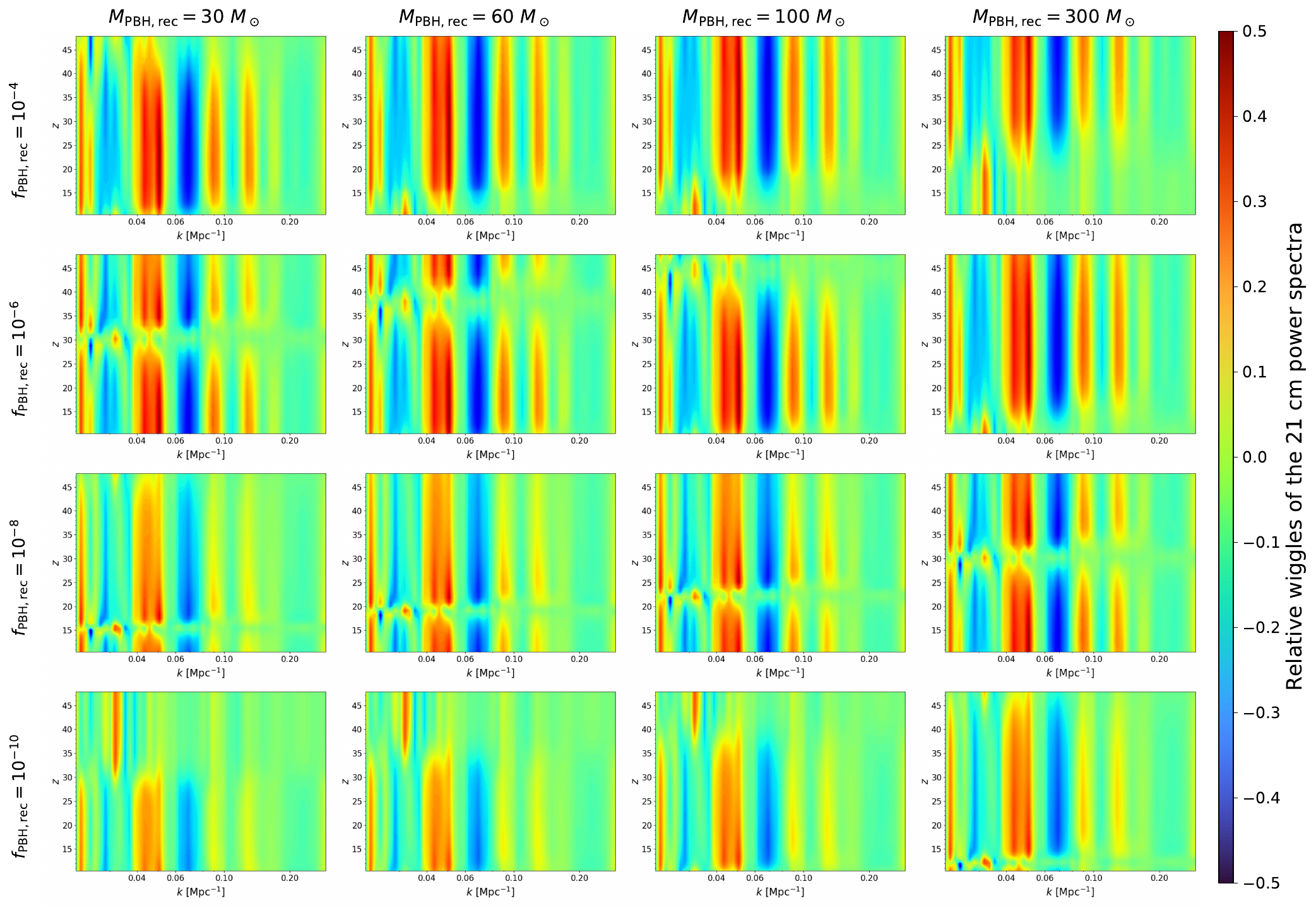}}
\caption{ The redshift evolution of the relative amplitude of the VAOs features in the 21 cm power spectrum at $z=20$, for different $M_{\rm PBH,rec}$ and $f_{\rm PBH,rec}$; Panels are arranged by PBHs parameters: from left to right $M_{\rm PBH,rec}=\{30,60,100,300\}M_{\odot}$; from bottom to top $f_{\rm PBH,rec}=\{10^{-10},10^{-8},10^{-6},10^{-4}\}$. }
\label{fig:contourf_redshift}
\end{figure*}

Fig. \ref{fig:contourf_redshift} shows the redshift evolution of the relative amplitude of VAOs wiggles in the 21 cm power spectrum, $[\Delta_{\rm 21}^2-\Delta_{\rm 21,poly}^2]/\Delta_{\rm 21,poly}^2$, arranged in a $4\times4$ matrix. From left to right, $M_{\rm PBH,rec}=\{10,30,100,300\}M_{\odot}$; from bottom to top, $f_{\rm PBH,rec}=\{10^{-10}, 10^{-8},10^{-6},10^{-4}\}$. We find that for given $M_{\rm PBH,rec}$ and $f_{\rm PBH,rec}$, it is not able to fully experience the five stages between $10 \le z \le 50$. For example, for $30~M_\odot \le M_{\rm PBH,rec}\le 300~M_\odot$, if $f_{\rm PBH,rec}\sim 10^{-10}$ there are only stage a) and b); while if $f_{\rm PBH,rec}\sim 10^{-8}$ there are only stage b), c) and d); if $f_{\rm PBH,rec}$ is as high as $\sim 10^{-4}$, then stage d) starts even as early as $z\sim 50$. 

The redshift for stage a) and b) transition, $z_\alpha$, and the redshift for stage c) and d) transition, $z_{\rm heat}$, can be described by a formula
\begin{equation}
    z_{\rm \alpha/heat} = p_{0} +p_{1} \log\left(\frac{M_{\rm PBH,rec}}{200M_{\odot}}\right) + p_{2} \log\left(\frac{f_{\rm PBH,rec}}{10^{-8}}\right), 
\label{eq:z_alpha_heat}
\end{equation}
and parameters are given in Table \ref{tab:transfits}. When $z \gtrsim z_\alpha$, there are no VAOs features; when $z_\alpha \le z \le z_{\rm heat}$, VAOs features arise from Ly$\alpha$ scattering; when $z \gtrsim z_{\rm heat}$, VAOs features arise from X-ray heating.

\begin{table}[t]
\centering
\caption{Values of parameters in Eq.~(\ref{eq:z_alpha_heat}), valid for $M_{\rm PBH,rec}\in[10,300]\,M_\odot$ and $f_{\rm PBH,rec}\in[10^{-14},10^{-2}]$
}
\label{tab:transfits}
\setlength{\tabcolsep}{15pt}
\begin{tabular}{lccc}
\hline
 & $p_{0}$ & $p_{1}$ & $p_{2}$ \\
\hline
$z_{\alpha}$   & $48.78$ & $11.57$ & $5.47$ \\
$z_{\rm heat}$ & $27.34$ & $18.29$ & $7.19$ \\
\hline
\end{tabular}
\end{table}

\section{The radio background from PBHs}\label{sec:radio-bg}

\subsection{
Radio Emission from PBHs
}

In addition to producing X-ray radiation via accretion, astrophysical black holes may also generate strong radio emission through jet mechanisms. If PBHs also accumulate angular momentum via accretion process and launch powerful jets in Dark Ages, their radio emission may build a radio background comparable to CMB. This  would enhance the 21 cm absorption strength.  Here we simply assume that 
the jet lasts a fraction $f_{\rm duty,jet}$ of the PBHs lifetime, the jet power is a fraction $\eta_{\rm jet}$ of the accretion power $L_{\rm acc}=\dot{M}_{\rm B}c^2=\dot{m}L_{\rm Edd}$, 
and the cumulative radio luminosity, defined as $\nu_{\rm 5}L_{\rm \nu_{5}}$, is a fraction $\eta_{\rm R}$ of the jet power, where $\nu_{\rm 5}$ is the frequency of 5 GHz. We assume the radio spectrum has a power-law form with a constant spectrum index:
\begin{align}
    L_{\rm R}(\nu,\dot{m},M_{\rm PBH})&=L_{\nu_{5}}\left(\frac{\nu}{\nu_{5}}\right)^{-\alpha_{\rm R}} \nonumber \\
    &=\frac{\eta_{\rm R}\eta_{\rm jet} \dot{m}L_{\rm Edd}}{\nu_{5}} \left(\frac{\nu}{\nu_{5}}\right)^{-\alpha_{\rm R}},
\end{align}
and adopt $\alpha_{\rm R}=0.75$. 

The specific intensity of radio background from PBHs is
\begin{equation}
    J_{\rm R}(\nu,\vec{r},z)= \frac{(1+z)^3}{4\pi} \int_{0}^{\infty} 
\tilde{\epsilon}'_{\rm R}(\nu',R'_{\rm c},\vec{r})\frac{dR'_{\rm c}}{1+z'},
\end{equation}
where the mean emissivity of the shell at radius $R'_{\rm c}$ from $\vec{r}$
\begin{align}
&\tilde{\epsilon}'_{\rm R}(\nu',R'_{\rm c},\vec{r}) =f_{\rm duty,jet} n_{\rm PBH} \times \nonumber \\
&\frac{1}{4\pi}\int_{4\pi} L_{\rm R}(\nu',\dot{m}(R'_{\rm c},\vec{\theta}),M_{\rm PBH}(R'_{\rm c},\vec{r}))d\Omega(\vec{\theta}).
\end{align}
The brightness temperature of the radio background from PBHs at the 21 cm frequency is
\begin{equation}
    T_{\rm R}(\nu_{21})=\frac{c^2}{2k_{\rm B}\nu_{21}^2}J_{\rm R}(\nu_{21}),
\end{equation}
where $\nu_{21}$ is the frequency of 21 cm line. 
This radio background is added to the $T_{\rm CMB}$ in Eq. (\ref{eq:deltaTb}) to calculate the new 21 cm signal.

\begin{figure}
\centering
\subfigure{\includegraphics[width=0.45\textwidth]{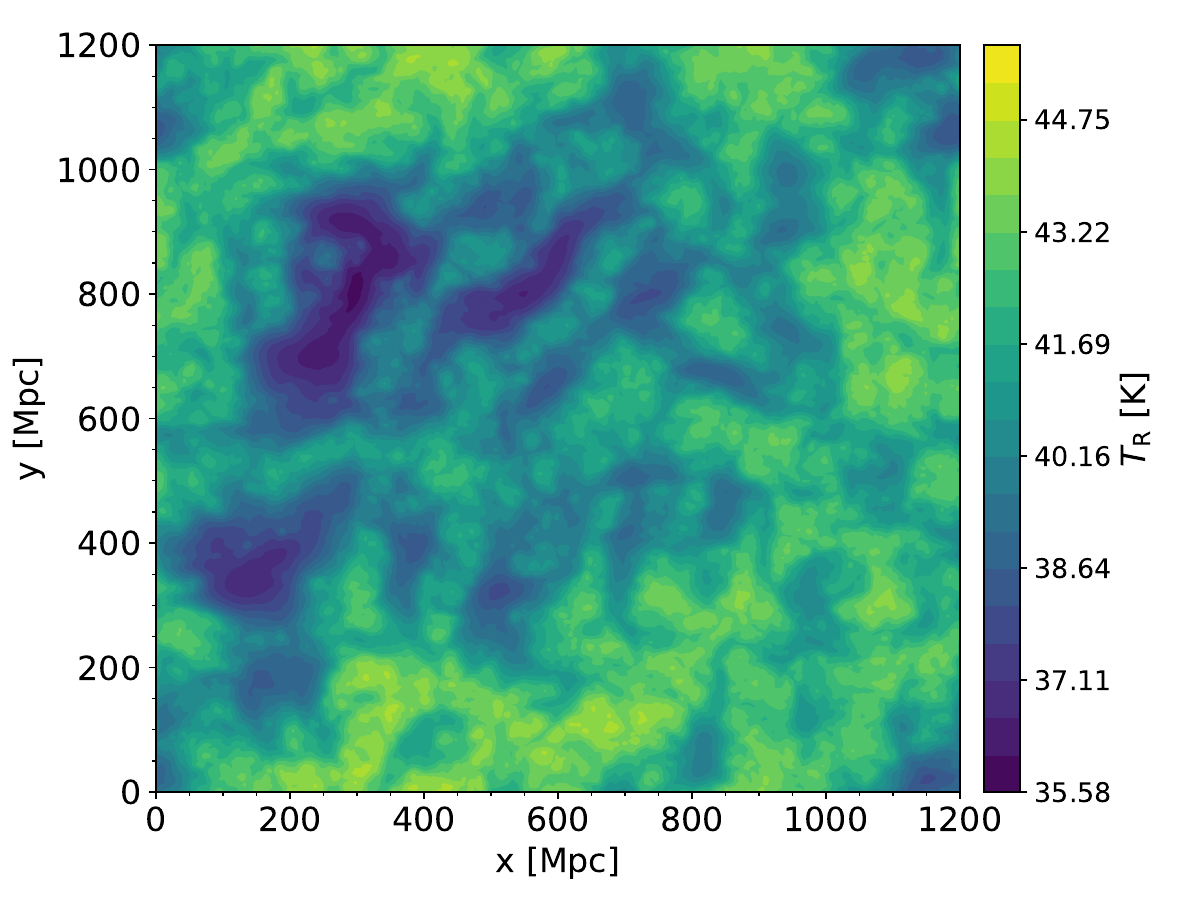}}
\subfigure{\includegraphics[width=0.45\textwidth]{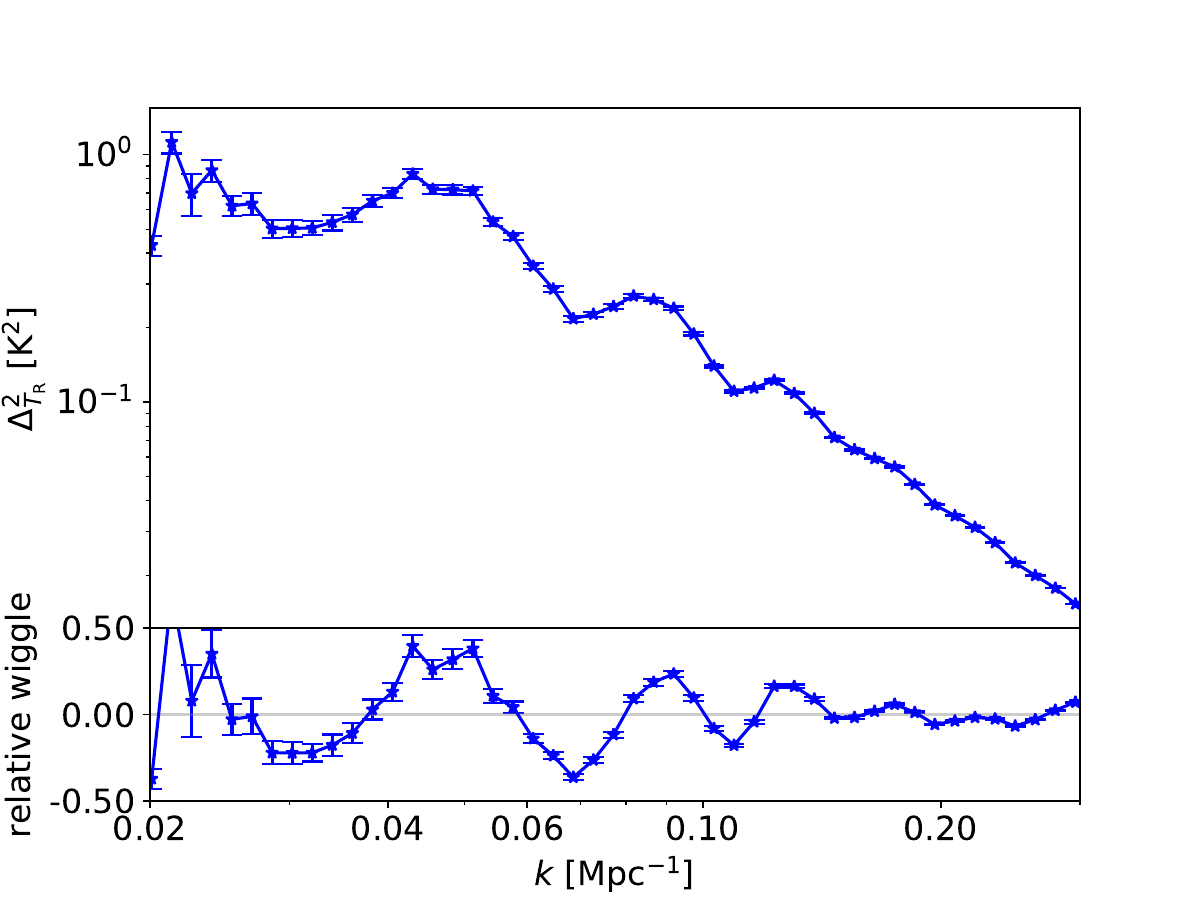}}
\caption{$Top$: A slice of the spatial distribution of the 1.4 GHz radio background at $z=20$, produced by PBHs with $M_{\rm PBH,rec}=200~M_{\odot}$,  $f_{\rm PBH,rec}=10^{-8}$, and $f_{\rm duty,jet}\eta_{\rm jet}\eta_{\rm R}=10^{-2}$.
$Bottom$: The power spectrum of the radio background, $\Delta_{T_{\rm R}}^2(k)$ (top sub-panel),   
and the relative amplitude of the VAOs wiggles (bottom sub-panel).  
}
\label{fig:T_CRB_D_P}
\end{figure}

The top panel of Fig. \ref{fig:T_CRB_D_P} shows a slice of  $T_{\rm R}$ at $z=20$, for $M_{\rm PBH,rec}=200~M_{\odot}$, $f_{\rm PBH,rec}=10^{-8}$, and $f_{\rm duty,jet}\eta_{\rm R}\eta_{\rm jet}=10^{-2}$. 
Although different from X-ray, the IGM is transparent to radio emission, $T_{\rm R}$ still exhibits remarkable spatial variations. Regions with higher accretion rates have  stronger radio background.  
The bottom panel of Fig. \ref{fig:T_CRB_D_P} presents the power spectrum of $T_{\rm R}$ and the corresponding relative amplitude of VAOs wiggles. Since radio photons are barely absorbed, their mean free path is much larger than X-ray radiation, and even much larger the Hubble scale. As a result, compared to the power spectrum of relative streaming velocities (Fig. \ref{fig:v_bc}), the radio background becomes smooth over a Hubble scale, leading to a suppression of power spectrum at $k>k_{\rm Hubble}$, where $k_{\rm Hubble}=4\times 10^{-3}~\rm Mpc^{-1}$ is the wavenumber of Hubble scale. 
However, the large-scale relative wiggles remain remarkable.

 \subsection{Impacts on the 21 cm Power Spectrum and VAOs Features}

\begin{figure}
\centering
\subfigure{\includegraphics[width=0.45\textwidth]{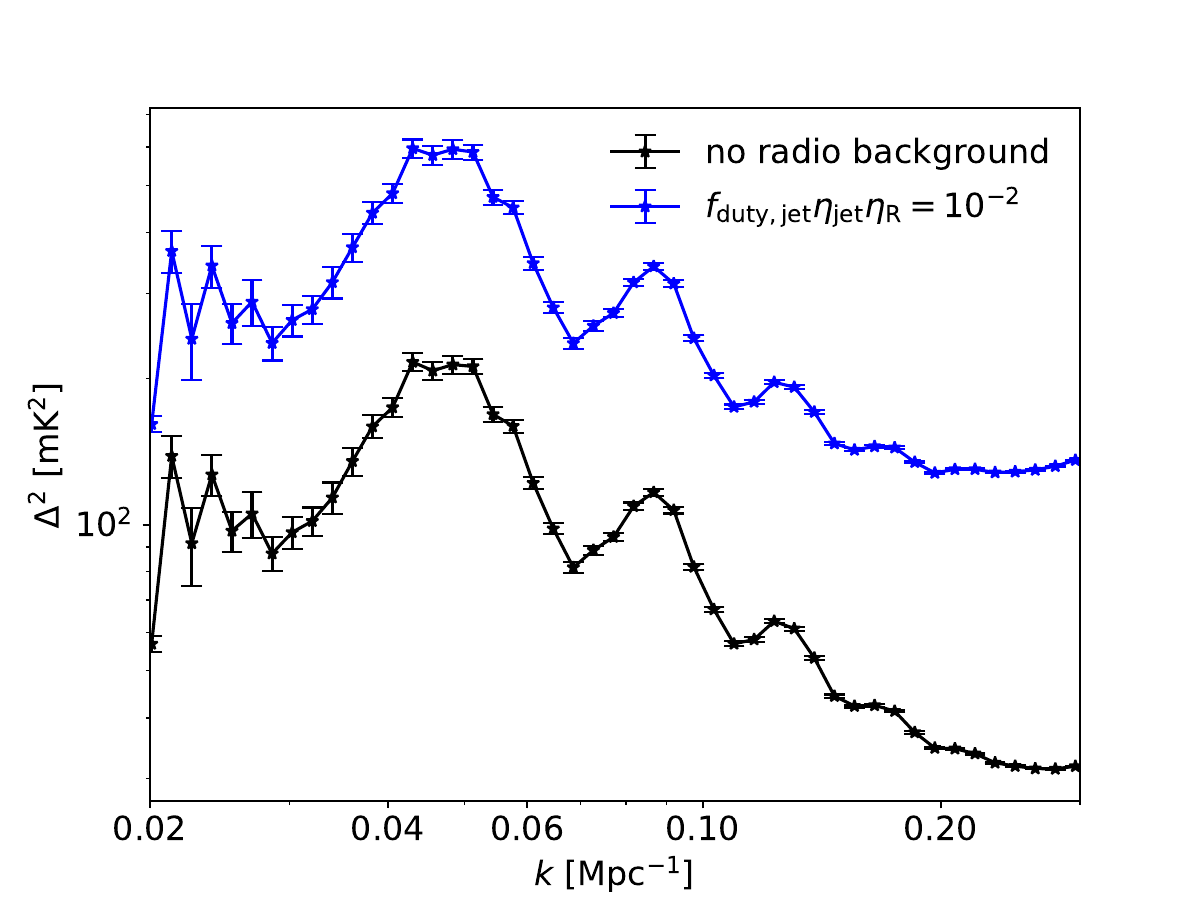}}
\caption{The 21 cm power spectrum  at $z=20$, for $M_{\rm PBH,rec}=200~M_{\odot}$ and $f_{\rm PBH,rec}=10^{-8}$, in the absence/presence of radio background from PBHs. 
} 
\label{fig:P_withCRB_and_noCRB}
\end{figure}

If PBHs generate a sufficiently strong radio background such that $T_{\rm R}\gtrsim T_{\rm CMB}$, then for 21 cm signal the global absorption will be deeper and the power spectrum will be stronger, while its shape remains unchanged, see Fig. \ref{fig:P_withCRB_and_noCRB}, where the radio background boosts the power  spectrum coherently across nearly all $k$-modes. 
This is because, although according to Eq. (\ref{eq:deltaTb}), radio background and X-ray heating have opposite effects on the 21 cm signal, the radio background from PBHs is spatially smooth on cosmological scales, therefore it primarily rescales the amplitude of the $\delta T_b$ at all scales. The shape of the power spectrum is not distorted and  the VAOs features are preserved.

\begin{figure}
\centering
\subfigure{\includegraphics[width=0.45\textwidth]{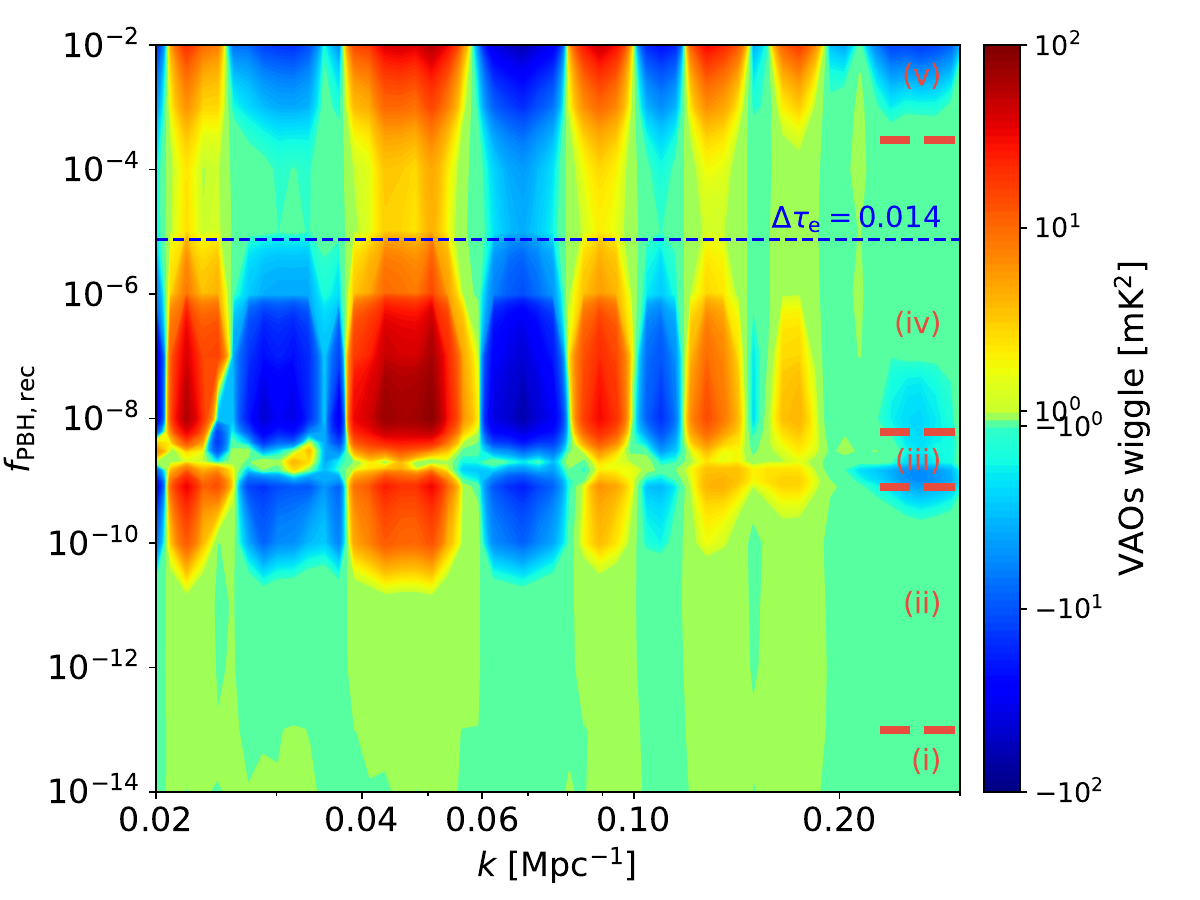}}
\caption{The relative amplitude of the VAOs features in the 21 cm power spectra at $z=20$,  as a function $f_{\rm PBH,rec}$, assuming $M_{\rm PBH,rec}=200~M_{\odot}$. 
Compared with Fig. \ref{fig:contourf_Pk_f_PBH}, here a fraction of PBHs can launch jet and build a strong radio background in Dark Ages, with ${f_{\rm duty,jet}\eta_{\rm jet}\eta_{\rm R}}=10^{-4}$. Blue dashed line is the same to Fig. \ref{fig:contourf_Pk_f_PBH}.
}
\label{fig:contourf_Pk_f_PBH_withCRB}
\end{figure}

More importantly, when $f_{\rm PBH,rec}$ is high enough and the IGM is heated to $T_{\rm K}\gg T_{\rm CMB}$, the VAOs features will become weak or even completely disappear, and the shape of 21 cm power spectrum traces the matter density field.
However, if PBHs also produce a strong radio background, then it is possible that $T_{\rm K} \gg T_{\rm CMB}$ but $T_{\rm K}$ is smaller or comparable to $T_{\rm CMB}+T_{\rm R}$, then the VAOs features survive. Fig. \ref{fig:contourf_Pk_f_PBH_withCRB} shows the relative amplitude of VAOs wiggles in the 21 cm power spectrum at $z=20$, for $M_{\rm PBH,rec}=200~M_{\odot}$ and ${f_{\rm duty,jet}\eta_{\rm jet}\eta_{\rm R}}=10^{-4}$. Compared with Fig. {\ref{fig:contourf_Pk_f_PBH}}, even for ${f_{\rm duty,jet}\eta_{\rm jet}\eta_{\rm R}}$ as small as $10^{-4}$, the two panels diverge once $f_{\rm PBH,rec}\gtrsim 10^{-4}$: in the absence of a radio background the VAOs wiggles vanishes, whereas with a radio background they remain clearly visible.

Both X-ray heating and radio background increase with $f_{\rm PBH,rec}$, to overcome the heating and preserve the VAOs features, generally it requires ${f_{\rm duty,jet}\eta_{\rm jet}\eta_{\rm R}} \gtrsim 10^{-5}$, which is more stringent than the fundamental plane for the radio luminosity and X-ray luminosity relation derived from observations of supermassive black holes \cite{Merloni_2003}.

\section{Detectability}\label{sec:detectability}

We use the {\tt 21cmSense} \citep{Murray2024} to estimate the uncertainties of the VAOs features observed by SKA-low AA* and a hypothetic lunar surface-based low-frequency interferometer array located at the farside of the Moon. 
For both arrays we adopt the tracking observation mode, focusing on a deep field as the primary beam.
For SKA-low AA*, we adopt the layout \texttt{ LOW\_SUBSTATION\_18M\_INNER\_R1KM\_AASTAR}, using the stations within 1 km radius. Since the VAOs features are large-scale signal, we group the antennae in each station (38 m diameter) into substations with diameter 18 m \cite{trott_2024_16951143}. 
This is to enlarge the field-of-view as possible, because the VAOs features are large-scale signal. For $k\sim 0.05$ Mpc$^{-1}$, if $z=20$, it corresponds angular size $\sim 38'$. 
So there are 856 substations in total. We adopt 180 observational days and each day has 6 h integration hours.

Compared with the ground-based telescopes on Earth, there is no ionosphere absorption/distortion at the farside of the Moon. Therefore such an array has the ability to detect the signal from higher redshifts \cite{2021ExA....51.1641K}. 
For the lunar surface-based array, we simply assume that there are 700 stations uniformly distributed within a circle with radius 500 m, and each station has a diameter of 18 m. The total integration time is assumed to be 9800 hours.

\begin{figure}
    \centering   
    {
\subfigure{\includegraphics[width=0.45\textwidth]{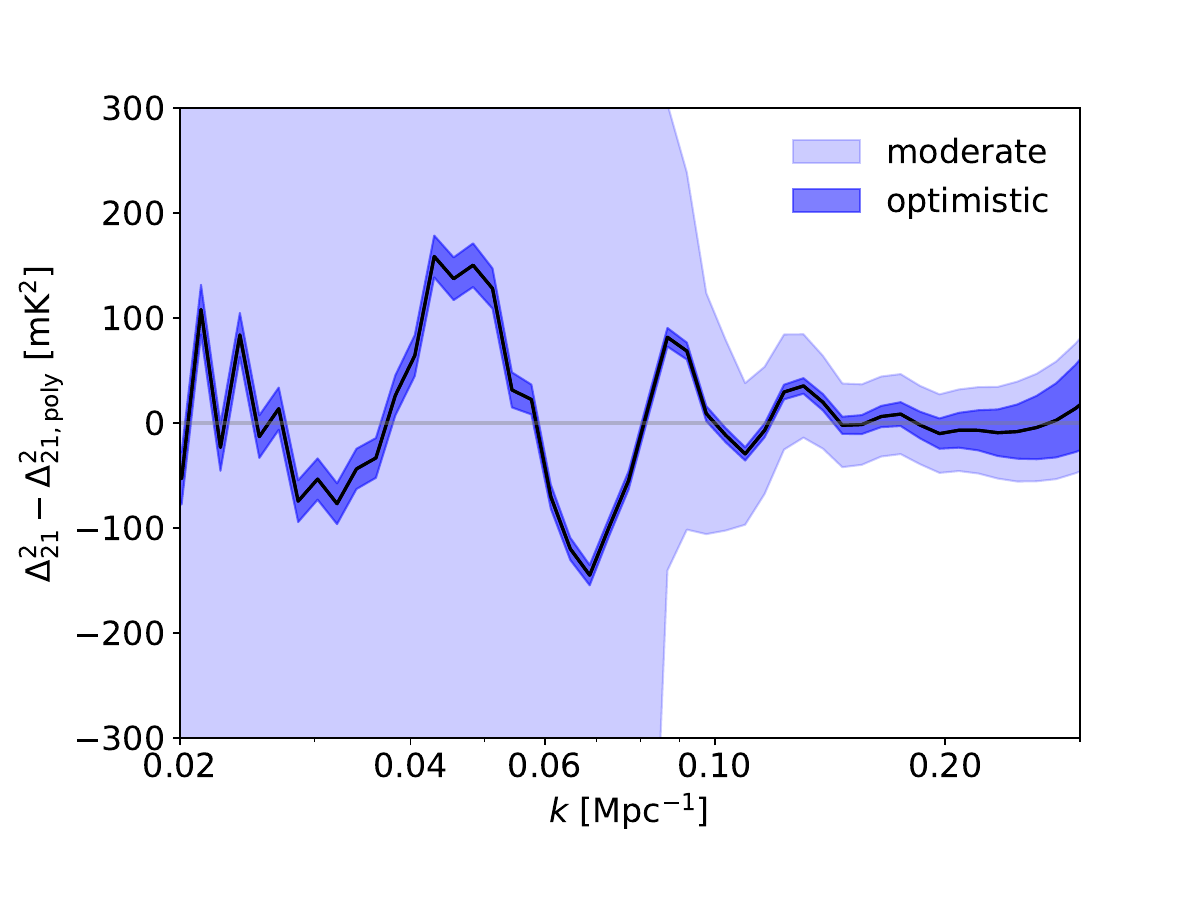}} 
\subfigure{\includegraphics[width=0.45\textwidth]{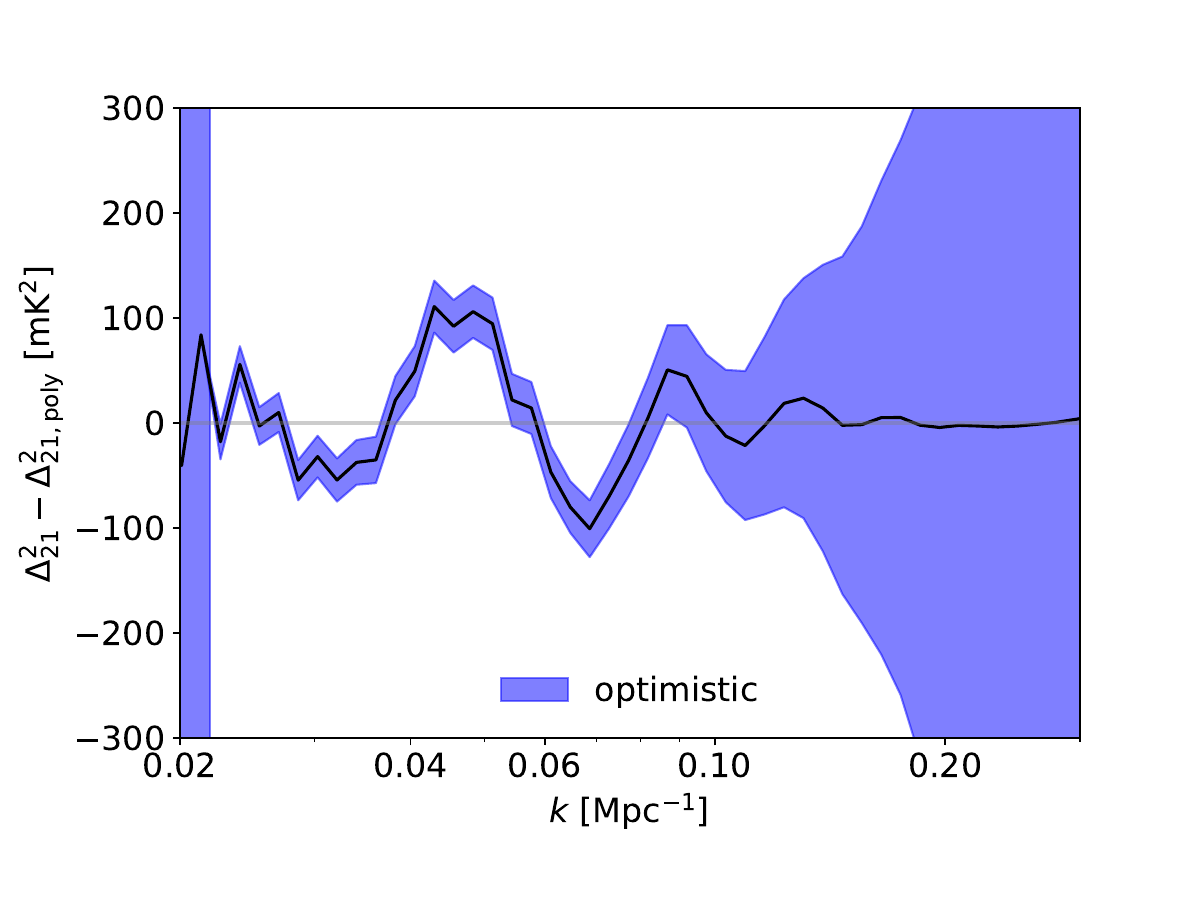}} 
    \caption{
    The net VAOs wiggles in the 21 cm power spectrum and the predicted 1$\sigma$ observational uncertainties, for $M_{\rm PBH,rec}=200~M_\odot$ and $f_{\rm PBH,rec}=10^{-8}$.    
    {\it Top:} The signal is from $z=20$ and the uncertainties are for SKA-low AA* with moderate (lighter region) and optimistic (darker region) foreground removal mode respectively. {\it Bottom:} The signal is  from $z=40$ and the  uncertainties are for a hypothetic lunar surface-based interferometer array with optimistic foreground removal mode. 
    \label{fig:21cmsense}
   }
    }
\end{figure}

The net VAOs wiggles, together with the predicted observational  uncertainties, are plotted in Fig. \ref{fig:21cmsense}. We see that, in principle, these VAOs features arise from PBHs accretion are detectable for reasonable setups of the upcoming or proposed instruments.

\section{Summary \& Discussion}\label{sec:summary}

\subsection{Summary}

In this paper, we investigated the impacts of relic supersonic DM-baryon relative streaming velocities on PBHs accretion and the 21 cm signal in Dark Ages. We adopted the scenario that each PBH can attract the surrounding DM particles to form an extended DM halo around it, then the DM halo boosts the accretion rate of central PBH compared with a naked PBH.  We generated the density filed and relative streaming velocity field, and computed the thermal evolution of the IGM and the 21 cm signal under the influence of X-ray radiation from PBHs accretion process. We found that: 
\begin{itemize}
\item In Dark Ages, the relative streaming velocities are strong enough to 
significantly reduce the PBHs accretion rate, imprinting their large-scale spatial pattern (the oscillatory wiggles in Fourier space) in the spatial distribution of accretion rates and the corresponding X-ray radiation. The Ly$\alpha$ coupling and gas heating induced by X-ray radiation also have this pattern, and finally generate VAOs features in the power spectrum of the 21 cm signal in Dark Ages. The relative amplitude of the first peak at $k\approx 0.05$ Mpc$^{-1}$ reaches up to $\sim30\%$. Since in Dark Ages, the astrophysical luminous objects have not yet formed, such VAOs features provide a potential chance to constraint the properties of PBHs.

\item A small PBHs fraction can generate the VAOs features in the 21 cm power spectrum. For example, if $M_{\rm PBH,rec}=200~M_\odot$, VAOs features appear at $z\sim 20$ even $f_{\rm PBH,rec}$ is as small as $\sim 10^{-13}$.

\item The redshift evolution of the VAOs features has five stages. In stage a) the influence of PBHs is negligible, there are no VAOs features in the 21 cm power spectrum. In stage b) the Ly$\alpha$ scattering becomes efficient and the VAOs features are generated by inhomogeneous Ly$\alpha$ scattering. In stage c) it is a transition stage, for which the VAOs features disappear temporarily. In stage d) the Ly$\alpha$ scattering has saturated, however the inhomogeneous X-ray heating generates VAOs features. In stage e) the gas is heavily heated, $T_{\rm S} \gg T_{\rm CMB}$, VAOs features disappear eventually. The redshift ranges of the five stages depend on $M_{\rm PBH,rec}$ and $f_{\rm PBH,rec}$.

\item At a given redshift, the dependence of VAOs features on $f_{\rm PBH,rec}$ and  $M_{\rm PBH,rec}$ can be classified into five phases.  Take $M_{\rm PBH,rec}=200~M_\odot$ and  $z=20$ as an example: If $f_{\rm PBH,rec}\lesssim 10^{-13}$, PBHs are not able to generate VAOs fearures in the 21 cm power spectrum at this redshift. If $10^{-13}\lesssim f_{\rm PBH,rec}\lesssim 2\times 10^{-9}$, the VAOs features are generated by inhomogeneous Ly$\alpha$ scattering. If $f_{\rm PBH,rec}\sim 2\times 10^{-9}$, the VAOs features disappear because Ly$\alpha$ scattering has been saturated but the X-ray heating is still negligible. If $2\times 10^{-9} \lesssim f_{\rm PBH,rec}\lesssim 10^{-3}$, inhomogeneous X-ray heating generates VAOs features in 21 cm power spectrum. Finally, if $f_{\rm PBH,rec} \gtrsim 10^{-3}$, the IGM is heavily heated and VAOs features disappear again.

\item When $T_{\rm S}\gg T_{\rm CMB}$, the 21 cm signal becomes independent of spin temperature and the VAOs features vanish. However, if PBHs also produce strong radio radiation and build an intense radio background $T_{\rm R}$ in Dark Ages, then the VAOs wiggles would be visible for longer, until $T_{\rm S} \gg T_{\rm CMB}+T_{\rm R}$. Such a radio background  boosts the amplitude of the 21 cm power spectrum, but the relative amplitude of VAOs wiggles almost remains the same.
\end{itemize}

\subsection{Discussion}\label{sec:discussion}

Strictly speaking, to investigate the VAOs effects on PBHs accretion, we should derive the statistics of the PBHs-baryon relative streaming velocities. So far as we know, there are no such scenario proposed in literatures, we therefore simply apply the statistics of the DM-baryon relative streaming velocities to PBHs. We believe this is a good approximation for the following reasons: 

(1) The VAOs effects are large-scale effects. The coherence scale is $\approx3$ Mpc, and the first peak of the oscillations is located at $k\approx 0.05$ Mpc$^{-1}$, corresponding to a correlation scale $125$ Mpc. Currently, there are no necessarily involved mechanism that can generate large-scale velocity offset between PBHs and DM.
At small scale ($\lesssim 3$ Mpc), even the motion of PBHs does not exactly trace the DM, it will just add some noise to the large-scale statistics.  

(2) Compared with the VAOs effects, PBHs formed much earlier. For example, for a PBH with mass $200~M_\odot$, it formed at $z\sim 10^{{11}}$ \cite{2020ARNPS..70..355C,2018arXiv181211011C}. After formation, PBHs behave as pressureless and
collisionless particles, like DM particles. In the linear regime at large-scale, PBHs must obey the same Euler equation as DM,  
    \begin{align}
        \dot{v}_{\rm DM}+H(z)v_{\rm DM}&=-\frac{\nabla\Phi}{a} \nonumber \\
        \dot{v}_{\rm PBH}+H(z)v_{\rm PBH}&=-\frac{\nabla\Phi}{a},
        \label{eq:v_PBHs_v_DM}
    \end{align}    
   where $v_{\rm DM}$ and $v_{\rm PBH}$ are velocities of DM and PBHs respectively; $\Phi$ is the total gravitational potential and $a$ is the scale factor. Actually, both PBHs and DM are in the same gravitational potential. Subtracting the second equation from the first equation, the solution of the  DM-PBHs relative velocity  is a decay mode and scales as $a^{-1}$ 
\citep{2019PhRvD.100h3528I,2010PhRvD..82h3520T}. That is, even  at $z\sim10^{11}$ when PBHs just formed, there is velocity offset between PBHs and DM, at $z\sim 10^3$ it has decayed by $\sim 8$ orders of magnitude. So the initial velocity offset between PBHs and DM, even do exist, will have negligible influence on the VAOs effects produced at the recombination era and later on.  

(3) There have been many papers that assume the PBHs-baryon relative velocities are close to the DM-baryon relative velocities, for example \cite{2008ApJ...680..829R,2017PhRvD..95d3534A,2022JCAP...12..016P}.

We have adopted the scenario that each PBH resides in a circum-PBH DM halo for our investigation. Such DM halos form because PBHs not only accrete the gas, but also attract DM in the environment.  
In this model, the host DM halo boosts the accretion rate of its central PBH. Alternatively, PBHs could be also naked and the accretion rate is much smaller for same PBHs mass. We checked that, using the naked PBHs scenario will not change our conclusion, except that to produce the same amplitude of VAOs wiggles, the required $f_{\rm PBH,rec}$ would be much larger. For example, for naked PBHs with $M_{\rm PBH,rec}=200 ~M_{\odot}$, VAOs features appear only if $f_{\rm PBH,rec} $ reaches $\sim 10^{-4}$, a value that is comparable to current observational upper limit constraints on the PBHs fraction in the DM \cite{2017PhRvD..95d3534A,2020PhRvR...2b3204S,Hektor2018PRD,2017PhRvD..96h3524P,2017JCAP...09..037R}.

We present a clarification about the ``PBHs fraction in the DM'' $f_{\rm PBH}$, and the ``PBHs abundance'' $\beta_{\rm PBH}$. $f_{\rm PBH}$ refers to the ratio between the PBH mass density and the DM density, this concept is generally used for PBHs in present-day Universe, or at least no earlier than the matter-dominated epoch (e.g. \cite{DAgostino_2023PhRvD.107d3032D}). 
So it is time-independent if the PBH mass growth via accretion is negligible. On the other hand, $\beta_{\rm PBH}$ refers to the fraction of PBHs in the total energy density, including the PBHs themselves, matter and radiation \citep{2010PhRvD..81j4019C,Carr_2016PhRvD..94h3504C,2021JPhG...48d3001G,Carr_2021RPPh...84k6902C,Carr_2024arXiv240605736C}.  This abundance is generally used when investigating the PBHs evolution from the formation time to the matter-dominated epoch. It evolves rapidly before the matter-radiation equality.
However, in the matter-dominated epoch,  $\beta_{\rm PBH}$ is also time-independent and the value is close to $f_{\rm PBH}$, just baryon is also included in the denominator for $\beta_{\rm PBH}$, therefore in some literatures $f_{\rm PBH}$ is also named PBHs abundance (e.g. \cite{2019PhRvD..99j3531W}).

The PBH is formed from the collapse of horizon-scale perturbation in the radiation-dominated epoch, so the initial mass is the enclosed mass within the Hubble radius at that time (Hubble mass). For PBHs with mass $M_{\rm PBH}^{\rm form}$ and density $\rho_{\rm PBH}^{\rm form}$ at the formation time, their fraction in the total energy density
\begin{align}
\beta_{\rm PBH}^{\rm form}(M_{\rm PBH}^{\rm form})&= \frac{\rho_{\rm PBH}^{\rm form}}{\rho_{\rm PBH}^{\rm form}+\rho_{\rm m}^{\rm form}+\rho_{\rm rad}^{\rm form}} \nonumber \\
&\approx \frac{\rho_{\rm PBH}^{\rm form}}{\rho_{\rm PBH}^{\rm form}+\rho_{\rm rad}^{\rm form}}
\end{align}
where $\rho_{\rm m}^{\rm form}$ and $\rho_{\rm rad}^{\rm form}$ are matter and radiation density at the PBH formation time respectively. The approximation holds because before the matter-radiation equality, the radiation density definitely dominates over the matter.

After formation, as the expansion of the Universe, the PBHs density scales with $a^{-3}$, while radiation scales with $a^{-4}$, therefore the abundance evolves as 
\begin{equation}
\beta_{\rm PBH}(M_{\rm PBH}^{\rm form},a)\approx \frac{   \rho_{\rm PBH}^{\rm form} }{ \rho_{\rm PBH}^{\rm form}+  (a^{-1}/a_{\rm form}^{-1})\rho^{\rm form}_{\rm rad} },
\end{equation}
where $a_{\rm form}$ is the scale factor at the formation time.
 $\beta_{\rm PBH}$ keeps increasing until it reaches 
\begin{equation}
\beta_{\rm PBH}^{\rm eq}(M_{\rm PBH}^{\rm form})\approx a_{\rm eq}/a_{\rm form} \beta_{\rm PBH}^{\rm form}(M_{\rm PBH}^{\rm form}),
\label{eq:beta_PBH_eq}
\end{equation}
where $a_{\rm eq}$ is the scale factor at the matter-radiation equality.

The Universe enters into the matter-dominated epoch after the matter-radiation equality, $\beta_{\rm PBH}$ no longer changes if the accretion is ignored, so one can define a PBHs fraction in the DM according to  $\beta_{\rm PBH}^{\rm eq}$ \citep{Carr_2016PhRvD..94h3504C},
\begin{align}
f_{\rm PBH}(M_{\rm PBH}^{\rm form})&=\frac{ \rho^{\rm eq}_{\rm PBH}(M_{\rm PBH}^{\rm form}) }{\rho^{\rm eq}_{\rm DM}} \nonumber \\
&=\frac{\beta^{\rm eq}_{\rm PBH}(M_{\rm PBH}^{\rm form})(\rho^{\rm eq}_{\rm m}+\rho^{\rm eq}_{\rm rad})}{(\rho^{\rm eq}_{\rm m}-\rho^{\rm eq}_{\rm b})} \nonumber \\
&=\frac{\beta_{\rm PBH}^{\rm eq}(M_{\rm PBH}^{\rm form})(2\rho^{\rm eq}_{\rm m})}{(\rho^{\rm eq}_{\rm m}-\rho^{\rm eq}_{\rm b})} \nonumber \\
&=2.4\beta_{\rm PBH}^{\rm eq}(M_{\rm PBH}^{\rm form}) ,
\end{align}
where $\rho_{\rm PBH}^{\rm eq}$, $\rho_{\rm DM}^{\rm eq}$, $\rho^{\rm eq}_{\rm m}$, $\rho^{\rm eq}_{\rm rad}$ and $\rho^{\rm eq}_{\rm b}$ are densities for PBH, DM, matter (DM+baryon), radiation and baryon respectively at the matter-radiation equality, and \citep{2020A&A...641A...6P}
\begin{equation}
\frac{2\rho_{\rm m}^{\rm eq}}{\rho_{\rm m}^{\rm eq}-\rho_{\rm b}^{\rm eq}}=\frac{2\Omega_{\rm m}}{\Omega_{\rm m}-\Omega_{\rm b}}=2.4.
\end{equation}

From the matter-radiation equality to the recombination era, this fraction just changes negligibly, therefore can be used as the initial condition for models describing the PBH evolution after the recombination era, as this paper.

Since $f_{\rm PBH}$ must be $\lesssim 1$, then $\beta_{\rm PBH}^{\rm eq}$ must be $\lesssim 0.4$. For the PBH formed much earlier than the matter-radiation equality, the initial fraction in the total energy must be very small \citep{Green_2024NuPhB100316494G}. For example, for the fiducial PBH mass $200~M_\odot$ in our paper, the formation redshift is $\sim 10^{11}$, then according to Eq. (\ref{eq:beta_PBH_eq}) the initial abundance must be $\lesssim  10^{-8}$.  This provides a very intuitive estimation for the PBH formation and evolution model.

The above derivation is applicable for the monochromatic mass distribution. If PBHs can form at different time with different initial mass, then $\beta_{\rm PBH}^{\rm eq}/M_{\rm PBH}^{\rm form}$ represents the mass spectrum, the abundance for all PBHs is then \citep{Carr_2016PhRvD..94h3504C}
\begin{equation}
\beta_{\rm PBH}^{\rm eq}=\int_{M_{\rm eva}}^{M_{\rm eq}} \frac{ \beta_{\rm PBH}^{\rm eq}(M_{\rm PBH}^{\rm form})}{M_{\rm PBH}^{\rm form}} dM_{\rm PBH}^{\rm form},
\end{equation}
where $M_{\rm eva}$ is the critical mass for Hawking evaporation and $M_{\rm eq}$ is the Hubble mass at matter-radiation equality.

\begin{acknowledgments}
We thank the anonymous referee very much for the constructive suggestions that help to largely improve our paper. This work is supported by the NSFC International (Regional) Cooperation and Exchange Project No. 12361141814, the National Key R\&D Program of China No. 2022YFF0504300, the National SKA Program of China Nos. 2020SKA0110402 \& 2020SKA0110401, the China's Space Origins Exploration Program Nos. GJ11010405 \& GJ11010401, and the Project Supported by the Specialized Research Fund for State Key Laboratory of Radio Astronomy and Technology. It partially uses the computing resources of the National Supercomputing Center in Tianjin.
\end{acknowledgments}

\bibliography{ms}

\end{document}